\documentclass[cernpreprint,atlasdraft=false,texlive=2016,USenglish,texmf]{atlasdoc}
 
\usepackage[subfigure]{atlaspackage}
\usepackage[backref=false]{atlasbiblatex}
 
\usepackage[BSM,jetetmiss]{atlasphysics}
 
\graphicspath{{logos/}{figures/}}
 
\usepackage{ANA-SUSY-2018-36-PAPER-defs}
\usepackage{multirow}

\AtlasTitle{Search for trilepton resonances from chargino and neutralino pair production in $\sqrt{s} = 13$~$\text{Te\kern -0.1em V}$ $pp$ collisions with the ATLAS detector}
 
\AtlasVersion{2.6}
\PreprintIdNumber{CERN-EP-2020-201}
 
\AtlasJournalRef{Phys. Rev. D 103, (2021) 112003}
\AtlasDOI{10.1103/PhysRevD.103.112003}

\AtlasAbstract{
 
A search is performed for the electroweak pair production of charginos and associated production of a chargino and neutralino, each of which decays through an $R$-parity-violating coupling into a lepton and a $W$, $Z$, or Higgs boson.
The trilepton invariant-mass spectrum is constructed from events with three or more leptons, targeting chargino decays that include an electron or muon and a leptonically decaying $Z$ boson.
The analyzed dataset corresponds to an integrated luminosity of 139~\ifb\ of proton--proton collision data produced by the Large Hadron Collider at a center-of-mass energy of $\sqs = 13$~\TeV and collected by the ATLAS experiment between 2015 and 2018.
The data are found to be consistent with predictions from the Standard Model.
The results are interpreted as limits at 95\% confidence level on model-independent cross sections for processes beyond the Standard Model.
Limits are also set on the production of charginos and neutralinos for a Minimal Supersymmetric Standard Model with an approximate \BL symmetry.
Charginos and neutralinos with masses between 100~\GeV and 1100~\GeV are excluded
depending on the assumed decay branching fractions into a lepton (electron, muon, or $\tau$-lepton) plus a boson ($W$, $Z$, or Higgs).
 
}
 
\author{The ATLAS Collaboration}
 
\AtlasRefCode{SUSY-2018-36}
 
\AtlasDate{\today}
 
\AtlasJournal{Phys. Rev. D}
 
\hypersetup{pdftitle={ATLAS document},pdfauthor={The ATLAS Collaboration}}
 
\begin{document}
 
\maketitle

\section{Introduction}
\label{sec:intro}
The extension of the Standard Model (SM) of particle physics with supersymmetry (SUSY)~\cite{Golfand:1971iw,Volkov:1973ix,Wess:1974tw,Wess:1974jb,Ferrara:1974pu,Salam:1974ig}
can introduce processes that violate baryon number ($B$) and lepton number ($L$) conservation, for instance proton decay.
As such processes have not been observed, it is common to introduce an \textit{ad hoc} requirement to
conserve $R$-parity~\cite{Farrar:1978xj}, where the $R$-parity of a particle is defined as $R=(-1)^{3(B-L)+2s}$.
Here the $B$, $L$, and $s$ are the baryon number, lepton number, and spin of the particle, respectively.
All SM particles have $R=1$ and their SUSY partners have $R=-1$.
$R$-parity conservation (RPC) therefore requires the lightest SUSY particle (LSP) to be stable.
In RPC scenarios, a stable LSP must necessarily be neutral in electric and color charge to be compatible with astrophysical data~\cite{Ellis:1983ew,Pospelov:2006sc}.
 
Theories predicting $R$-parity violation (RPV)~\cite{Dreiner:1997uz,Barbier:2004ez} are viable if the interactions that violate \BL conservation have small couplings and violate only one of $B$ or $L$ at tree level, thus preventing rapid proton decay.
The benchmark model for this search is a Minimal Supersymmetric Standard Model (MSSM)~\cite{Fayet:1976et,Fayet:1977yc} extension that adds a gauged $U(1)_{\BL}$~\cite{Barger:2008wn,Everett:2009vy,Braun:2005ux,Deen:2016vyh,FileviezPerez:2008sx}
to the $SU(3)_\mathrm{C} \times SU(2)_\mathrm{L} \times U(1)_\mathrm{Y}$ of the SM and includes three generations of right-handed neutrino supermultiplets.
Any one of the right-handed sneutrinos has the correct quantum numbers to spontaneously break the \BL symmetry,
and its vacuum expectation value (VEV) introduces $L$ violation only at tree level~\cite{Deen:2016vyh}.
The size of the RPV coupling is directly related to the right-handed sneutrino VEV, and therefore to the neutrino sector.
As a consequence the RPV coupling is kept small by the small values of the neutrino masses.
The LSP may decay into SM particles through the RPV coupling, which allows the LSP to have electric and color charges.
 
\begin{figure}[h]
\centering
\includegraphics[width=0.48\textwidth]{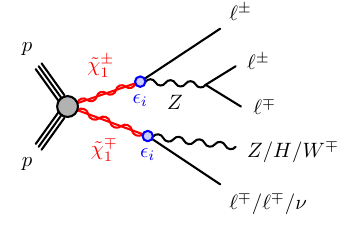}
\includegraphics[width=0.48\textwidth]{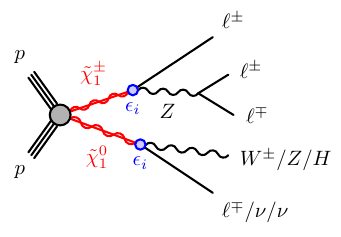}
\caption{Diagrams of (left) \CCsignal and (right) \CNsignal production with at least one \triLepDecay decay.
The $R$-parity-violating coupling $\epsilon_i$ allows prompt \chone decays into \Zl, \Hl, or \Wv and prompt \none decays into \Wl, \Zv, or \Hv.
}
\label{fig:feynman_diagram}
\end{figure}

The \BL RPV model predicts unique signatures~\cite{FileviezPerez:2012mj, Perez:2013kla} that are forbidden if $R$-parity conservation is assumed.
In a set of simulations~\cite{Ovrut:2014rba,Ovrut:2015uea} the MSSM parameters were scanned and the exact physical sparticle spectrum was calculated for each simulated point.
It was seen~\cite{Dumitru2019,dumitru2018rparity} that two likely LSP candidates with moderate production cross sections at the Large Hadron Collider (LHC) are the wino-type chargino (\chone) and wino-type neutralino (\none), the SUSY partners of the electroweak gauge fields of the $W$ bosons.
Both LSP candidates were found to be nearly mass degenerate with one another for all simulations and therefore both decay primarily via RPV couplings~\cite{dumitru2018rparity}.
The RPV coupling was also found by the simulations to be large enough that both the \chone and \none decay promptly~\cite{dumitru2018rparity}.
Therefore, this search targets prompt decays.
In this model the chargino may decay into a $Z$ boson and a charged lepton (\Zl), a Higgs boson and a charged lepton (\Hl), or a $W$ boson and a neutrino (\Wv),
while the neutralino may decay into \Wl, \Zv, or \Hv, as shown in Figure~\ref{fig:feynman_diagram}.
The \chone and \none branching fractions depend on tan\(\beta\), the ratio of the VEVs of the two Higgs fields, and the neutrino mass hierarchy.
For example, the branching fractions to electrons are predicted to be small in the normal hierarchy.
 
This paper presents a search for the electroweak pair production of two charginos (\CCsignal) or associated production of a chargino and neutralino (\CNsignal).
In contrast to RPC searches, there is no significant missing transverse momentum from an invisible LSP in the event, and all decay products can leave visible energy deposits in the detector.
A resonance search in the trilepton mass (\mZl) is performed in three orthogonal signal regions, all of which target events where the decay of at least one \chone forms a trilepton resonance.
One signal region requires four or more leptons and targets events where the second \chone or \none (denoted hereafter by\chono) decay can be fully reconstructed.
A second signal region also requires four or more leptons but targets decays of the second \chono that include one or more leptons and at least one neutrino.
A third signal region requires exactly three leptons, targeting decays of the second \chono that include no leptons.
 
Several SM processes with similar final-state particles can contribute to the signal regions, with the largest contributions from the \WZ, \ZZ, and \ttZ processes.
The expected yields of these processes are estimated using Monte Carlo (MC) simulation that is normalized to data in three highly populated control regions.
Additional event selections are applied to reject events from SM processes in the signal regions while maintaining a high selection efficiency for events from the target \CCsignal and \CNsignal models.
 
A scan over the possible \chone and \none branching fractions to both bosons and leptons is performed when setting model-specific limits.
Model-independent limits are also explored in narrow slices of the \mZl spectrum,
with no assumptions made on the \chono branching fractions or decay kinematics of a generic beyond-the-SM process.
 
Previous searches for the production of wino-type charginos and neutralinos in $R$-parity-conserving models
have targeted final states with three or more leptons via $W$ and $Z$ boson decays and found no significant excess in data over background expectations,
with the ATLAS~\cite{SUSY-2016-24,SUSY-2013-12} and CMS~\cite{CMS-SUS-16-039,CMS-SUS-17-004} collaborations setting limits on wino masses of up to 580~\GeV and 650~\GeV, respectively.
Searches have also been performed for trilepton resonances from heavy leptons in type-III seesaw scenarios by the ATLAS~\cite{EXOT-2014-07,EXOT-2014-08} and CMS~\cite{CMS-EXO-17-006} collaborations,
but none have attempted to fully reconstruct both decay chains of the charginos and neutralinos.
A previous search by ATLAS~\cite{SUSY-2016-29} for events from the \BL RPV model targeted by this analysis focused on the pair production of top squarks~\cite{Marshall:2014cwa}.
 
A brief overview of the ATLAS detector is given in \secref{detector},
and a description of the dataset and the MC simulation is presented in \secref{datamc}.
Details of the reconstruction of the events used in the search are presented in \secref{eventreco},
and the design of signal regions sensitive to the \BL RPV model is discussed in \secref{search}.
The description of the SM backgrounds and the strategy for their estimation are given in \secref{background},
followed by an explanation of the systematic uncertainties in \secref{systematics}.
The results of the search and their interpretation for various \BL RPV model scenarios are presented in \secref{result},
and the conclusions are given in \secref{conclusion}.
 
\section{ATLAS detector}
\label{sec:detector}
The ATLAS detector~\cite{PERF-2007-01} is a multipurpose particle detector with a nearly 4$\pi$ coverage in solid angle.\footnote{
ATLAS uses a right-handed coordinate system with its origin at the nominal interaction point (IP) in the center of the detector and the \(z\)-axis along the beam pipe.
The \(x\)-axis points from the IP to the center of the LHC ring, and the \(y\)-axis points upwards.
Cylindrical coordinates \((r,\phi)\) are used in the transverse plane, \(\phi\) being the azimuthal angle around the \(z\)-axis.
The pseudorapidity is defined in terms of the polar angle \(\theta\) as \(\eta = -\ln \tan(\theta/2)\),
and the rapidity $y$ is defined as $y= (1/2) \text{ln}[(E+p_z)/(E-p_z)]$, where $E$ is energy and $p_z$ is longitudinal momentum.
Angular distance is measured in units of \(\Delta R \equiv \sqrt{(\Delta\eta)^{2} + (\Delta\phi)^{2}}\).}
It is composed of an inner tracking system covering the pseudorapidity region $|\eta|<2.5$, electromagnetic and hadronic calorimeters covering $|\eta|<4.9$, and a muon spectrometer covering $|\eta|<2.7$.
 
The inner detector (ID) reconstructs tracks from charged particles using silicon pixel, silicon microstrip, and transition-radiation tracking detectors.
The innermost layer of the silicon pixel tracker, the insertable B-layer~\cite{ATLAS-TDR-19,PIX-2018-001}, was installed prior to 2015 at an average radial distance of 3.3~cm from the beamline to improve track reconstruction and the identification of jets initiated by heavy-flavor hadrons.
The ID is surrounded by a thin superconducting solenoid providing a \SI{2}{\tesla} axial magnetic field, facilitating the measurement of charged-particle momenta.
 
Beyond the solenoid is a high-granularity lead and liquid-argon (LAr) electromagnetic sampling calorimeter covering $|\eta|<3.2$.
Outside the electromagnetic calorimeter are two hadronic calorimeters; a steel and scintillator-tile sampling calorimeter covering $|\eta|<1.7$ and a copper and LAr endcap calorimeter covering $1.7<|\eta|<3.2$.
The most forward region of $3.1<|\eta|<4.9$ is covered by copper and LAr and tungsten and LAr calorimeters optimized for electromagnetic and hadronic measurements, respectively.
 
The muon spectrometer (MS) surrounds the calorimeters, identifying and measuring muon tracks through up to three layers of precision tracking and triggering chambers.
The MS is interleaved with a system of three superconducting air-core toroidal magnets with eight coils each, with a field integral between \SI{2.0}{\tesla\metre} and \SI{6.0}{\tesla\metre} across most of the detector.
 
The ATLAS trigger system consists of a hardware-based first-level (L1) trigger followed by a software-based high-level trigger (HLT)~\cite{TRIG-2016-01}.
The L1 and HLT trigger systems are designed to accept events at average rates of 100~kHz and 1~kHz, respectively.
Candidate electrons within $|\eta|<2.5$ are identified by the L1 trigger as compact electromagnetic energy deposits in the electromagnetic calorimeter, and by the HLT using additional fast track reconstruction~\cite{TRIG-2018-05}.
Candidate muons within $|\eta|<2.7$ are identified by the L1 trigger through a coincidence of MS trigger chamber layers and further selected by the HLT using fast reconstruction algorithms with input from the ID and MS.
 
 
\section{Data and Monte Carlo simulation}
\label{sec:datamc}
The analysis is performed using $pp$ collision data collected by the ATLAS experiment between the years 2015 and 2018.
The dataset corresponds to a total integrated luminosity of $139~\ifb$ after imposing data quality requirements~\cite{DAPR-2018-01}.
In this dataset there are, on average, approximately 34 simultaneous $pp$ collisions in each LHC proton bunch crossing.
 
Monte Carlo simulation is used to model the expected contributions of various SM processes and the \CCsignal and \CNsignal signal processes targeted by the search.
It is used to define and optimize the event selection criteria and to
estimate systematic uncertainties in the predictions of event yields.
The generators and parameters used in the MC simulation samples are given below and summarized in Table~\ref{tab:MC}.
The expected yields of SM processes are taken directly from MC simulation except for the dominant \WZ, \ZZ, and \ttZ backgrounds,
which are estimated from MC simulation that is normalized to data in dedicated control regions, as described in Section~\ref{sec:background:MC}.
The contribution from events with one or more misidentified or nonprompt (\fake) leptons is separately predicted using a data-driven method described in Section~\ref{sec:background:Fake}.
 
\begin{table}[t]
\footnotesize
\centering{
\caption{
Details of the MC simulation for each physics process, including the event generator used for matrix element calculation,
the generator used for the PS and hadronization, the PS parameter tunes, and the order in \alphas of the production cross-section calculations.
}
\label{tab:MC}
\begin{tabular}{lcccc}
\toprule
\multirow{2}{*}{Process} & \multirow{2}{*}{Event generator}    & PS and        & \multirow{2}{*}{PS tune} & \multirow{2}{*}{Cross section (in QCD)} \\
&                                     & hadronization &                          & \\
\hline
\hline
 
Diboson, triboson, (\Zjet)  & \SHERPAV{2.2}                 & \SHERPAV{2.2} & Default & \scriptsize{NLO (NNLO)} \\
\ttW, \ttZ, (Other top)     & \scriptsize{\MGMCatNLOV{2}}   & \PYTHIAV{8}   & A14     & \scriptsize{NLO (LO)} \\
\ttbar, (\tW), [\ttH{}]     & \POWHEGBOX~v2                 & \PYTHIAV{8}   & A14     & \scriptsize{NNLO+NNLL (NLO+NNLL) [NLO]} \\
Higgs: ggF, (VBF, $VH$)     & \POWHEGBOX~v2                 & \PYTHIAV{8}   & AZNLO  & \scriptsize{NNNLO (NNLO+NNLL)} \\
\CCsignal, \CNsignal        & \MADGRAPHV{2.6}               & \PYTHIAV{8}   &  A14    & \scriptsize{NLO+NLL} \\
 
\bottomrule
\end{tabular}
}
\end{table}
 
Diboson, triboson, and \Zjet samples~\cite{ATL-PHYS-PUB-2017-005,ATL-PHYS-PUB-2017-006} were simulated using the \SHERPAV{2.2}\cite{Bothmann:2019yzt} generator.
Triboson and most diboson processes were simulated with \SHERPAV{2.2.2} while \Zjet and semileptonically decaying diboson processes were simulated with \SHERPAV{2.2.1}.
The matrix element calculations were matched to the parton shower (PS) simulation using Catani--Seymour dipole factorization~\cite{Gleisberg:2008fv,Schumann:2007mg}.
The matching was performed separately for different jet multiplicities and merged into an inclusive sample using an improved Catani--Krauss--Kuhn--Webber (CKKW) matching procedure~\cite{Catani:2001cc,Hoeche:2009rj}
extended to next-to-leading-order (NLO) accuracy in QCD using the MEPS@NLO
prescription~\cite{Hoeche:2011fd,Hoeche:2012yf,Catani:2001cc,Hoeche:2009rj}.
The virtual QCD correction for matrix elements at NLO accuracy was provided by the \openloops\ library~\cite{Cascioli:2011va,Denner:2016kdg}.
The \nnpdfnnlo~\cite{Ball:2014uwa} set of parton distribution functions (PDFs) was used
together with a dedicated set of tuned PS parameters (tune) developed by the \sherpa authors~\cite{Schumann:2007mg}.
 
The \Zjet (diboson) samples were calculated for up to two (one) additional partons at NLO and up to four (three) additional partons at leading order (LO) in QCD,
and the triboson samples were calculated at NLO in QCD for the inclusive processes and at LO in QCD for up to two additional parton emissions.
Diboson samples include loop-induced and electroweak production.
The diboson and triboson samples do not include Higgs boson contributions.
The cross sections calculated by the event generators were used for all samples except for \Zjet, which was normalized to a next-to-next-to-leading-order (NNLO) cross-section prediction~\cite{Anastasiou:2003ds}.
 
The \ttbar~\cite{Frixione:2007nw}, \ttH~\cite{Hartanto:2015uka}, and \tW~\cite{Re:2010bp} process samples were simulated at NLO in QCD using the \POWHEGBOX~\cite{Nason:2004rx,Frixione:2007vw,Alioli:2010xd}~v2 generator and the \nnpdfnlo PDF set.
The matrix element calculations were interfaced with \PYTHIAV{8.230}~\cite{Sjostrand:2014zea} for the PS using the A14 tune~\cite{ATL-PHYS-PUB-2014-021} and the \nnpdftwo PDF set~\cite{Ball:2012cx}.
The \hdamp\ parameter\footnote{The \hdamp\ parameter controls the transverse momentum \pt\ of the first additional emission beyond the leading-order Feynman diagram in the PS and therefore regulates the  high-\pt\ emission against which the \ttbar\ system recoils.}
was set to be 1.5 times larger than the top-quark mass following optimization studies using data~\cite{ATL-PHYS-PUB-2016-020}.
The \ttbar\ inclusive production cross section was corrected to the theory prediction calculated at NNLO in QCD and included the resummation of next-to-next-to-leading logarithmic (NNLL) soft-gluon terms calculated with \textsc{Top++2.0}~\cite{Czakon:2011xx}.
The \tW inclusive production cross section was corrected to the theory prediction at NLO in QCD with NNLL corrections to the soft-gluon terms~\cite{Aliev:2010zk,Kant:2014oha}.
Both samples were generated in the five-flavor scheme, setting all quark masses to zero except for the top quark.
The diagram-removal strategy~\cite{Frixione:2008yi} was employed in the \tW sample to remove the interference with \ttbar\ production~\cite{ATL-PHYS-PUB-2016-020}.
 
Other top-quark production processes were simulated with the \MGMCatNLOV{v2}~\cite{Alwall:2014hca} generator at either NLO in QCD with the \nnpdfnlo PDF set or at LO in QCD using the \nnpdftwo PDF set.
They were interfaced with \PYTHIAV{8} using the A14 tune and the \nnpdftwo PDF set.
Generator versions \MGMCatNLOV{v2.3} and \PYTHIAV{8.212} were used for \tZ, \tWZ, \ttZ, \ttW, and \ttWZ processes,
while versions \MGMCatNLOV{v2.2} and \PYTHIAV{8.186} were used for \ttg, \ttWW, and four-top processes.
These top-quark processes were generated at LO in QCD with the exception of \ttZ, \ttW, and \tWZ, which were generated at NLO in QCD.
 
Higgs boson production via gluon--gluon fusion (ggF) was simulated at NNLO accuracy in QCD using
the \POWHEGBOXV{v2} NNLOPS program~\cite{Hamilton:2013fea} and interfaced with \PYTHIAV{8.212}
using the AZNLO~tune~\cite{STDM-2012-23} and PDF4LHC15 NNLO PDF set~\cite{Butterworth:2015oua}.
The MC prediction was normalized to the next-to-next-to-next-to-leading-order (NNNLO) cross section in
QCD plus electroweak corrections at NLO~\cite{Aglietti:2004nj,Actis:2008ug}.
 
Higgs boson production via vector-boson fusion (VBF) and Higgs boson production in association with a $W$ or $Z$ boson ($VH$) were generated
using \POWHEGBOXV{v2} and interfaced with \PYTHIAV{8.212} using the AZNLO tune and CTEQ6L1~\cite{Pumplin:2002vw} PDF set.
The \powheg predictions are accurate to NLO in QCD and were tuned to match calculations including effects due to finite heavy-quark masses and soft-gluon resummations up to NNLL.
The MC predictions were normalized to NNLO QCD cross-section calculations with NLO electroweak corrections~\cite{Ciccolini:2007jr,Willenbrock1991,Brein_2004,Ciccolini_2003}.
 
The \BL RPV \CCsignal and \CNsignal signal samples were produced using \MGMCatNLOV{v2.6} and the \nnpdftwo PDF set with up to two additional partons calculated at LO in QCD and interfaced with \PYTHIAV{8.230} using the A14 tune and \nnpdftwo PDF set.
The scale parameter for jet--parton CKKW-L matching was set to a quarter of the \chinoonepm/\ninoone mass.
Samples were generated at masses between 100~\GeV and 1500~\GeV in steps of 50~\GeV.
Signals with masses below 100~\GeV were not explored as they have been excluded by previous three-lepton searches for charginos and neutralinos~\cite{SUSY-2016-24,SUSY-2013-12,CMS-SUS-16-039,CMS-SUS-17-004,EXOT-2014-07,EXOT-2014-08,CMS-EXO-17-006}.
 
Signal events were generated with equal \chono branching fractions to each boson ($W$, $Z$, or
Higgs bosons where kinematically accessible) plus charged-lepton ($e$, $\mu$, or $\tau$-lepton) channel.
In order to explore different assumptions for the \chono branching fractions in the analysis,
simulated events are reweighted appropriately, assuming that the \chone and \none branching fractions change in the same way.
 
Generated signal events were filtered to have at least three leptons, two of which were associated with a $Z$ boson.
Hadronically decaying $\tau$-leptons were not considered by this three-lepton filter for the \CNsignal events, increasing the useful statistics of the MC sample.
The \chone were also required to decay via a $Z$ boson in the \CNsignal events to increase the number of events with a trilepton resonance.
The inclusive production cross sections were calculated assuming mass-degenerate, wino-like \chinoonepm and \ninoone, as predicted by the \BL RPV model~\cite{Dumitru2019},
and were calculated at NLO in QCD with next-to-leading-logarithmic (NLL) corrections to the soft-gluon terms~\cite{Beenakker:1999xh,Debove:2010kf,Fuks:2012qx,Fuks:2013vua,Fiaschi:2018hgm}.
The cross sections and their uncertainties were derived from an envelope of cross-section predictions using different PDF sets and factorization and renormalization scales~\cite{Borschensky:2014cia}.
The inclusive cross sections for \CCsignal (\CNsignal) production at a center-of-mass energy of $\sqs=13~\TeV$ range from $11.6\pm0.5~(22.7\pm1.0)$~pb for masses of 100~\GeV to $0.040\pm0.006~(0.080\pm0.013)$~fb for masses of 1500~\GeV.
 
The modeling of $c$- and $b$-hadron decays in samples generated with \POWHEGBOX or \MGMCatNLO was performed with \evtgen~1.2.0~\cite{Lange:2001uf}.
Events from all generators were propagated through a full simulation of the ATLAS detector~\cite{SOFT-2010-01} using \textsc{Geant4}~\cite{Agostinelli:2002hh}
to model the interactions of particles with the detector.
A parameterized simulation of the ATLAS calorimeter~\cite{SOFT-2010-01} was used for faster detector simulation of signal, \tW, and \ttH processes
and was found to be in agreement with the full simulation.
The effect of multiple interactions in the same and neighboring bunch crossings (pileup) was modeled by overlaying simulated minimum-bias collisions onto each hard-scattering event.
The minimum-bias events were generated with \PYTHIAV{8.210} using the A3 tune~\cite{ATL-PHYS-PUB-2016-017} and \nnpdftwo PDF set.
For each simulated hard-scatter process a separate MC sample is generated to reflect the conditions of the 2015+2016, the 2017, and the 2018 datasets.
The number of overlaid minimum-bias collisions is sampled for each event according to the distribution of the average number of interactions per bunch crossing measured in that dataset.
 
 
\section{Event reconstruction}
\label{sec:eventreco}
The data events used in the analysis were recorded during stable beam conditions at the LHC and were required to meet data quality criteria.
Data events were collected with triggers requiring at least a single electron or a single muon reconstructed by the trigger system,
with various lepton-\pt thresholds depending upon the relative quality (including isolation) of the trigger-level leptons~\cite{TRIG-2016-01}.
In the analysis, tighter quality and \pt requirements are applied to the fully reconstructed signal leptons, as described below,
to ensure the event selection is free from bias in the trigger reconstruction.
Each event for which the trigger was activated is required to have at least one electron (muon)
with a fully calibrated \pt above 27, 61, or 141~\GeV (27.3 or 52.5~\GeV),
with larger-\pt requirements corresponding to reduced lepton-quality requirements of the trigger.
For the 2015 data, the \pt requirement of the analysis for the loosest-quality electron trigger is lowered to 121~\GeV.
The single-lepton triggers are found to be more than 90\% efficient for the signal model with mass of 100~\GeV and more than 99\% efficient for signal models of mass 300~\GeV or higher.
 
Both the data and MC events are required to have at least one reconstructed vertex that is associated with two or more tracks of transverse momentum $\pt > 500~\MeV$.
The primary vertex of each event is selected as the vertex with the largest $\Sigma\pt^2$ of associated tracks~\cite{ATL-PHYS-PUB-2015-026}.
 
The primary objects considered by this analysis are electrons, muons, and jets.
Electron candidates are reconstructed from three-dimensional energy clusters in the electromagnetic calorimeter
that are matched to an ID track
and calibrated \insitu using $Z\rightarrow ee$ decays~\cite{EGAM-2018-01}.
Muon candidates in the detector are typically reconstructed from a combined fit of tracks formed in the MS and ID
and calibrated \insitu using $Z \rightarrow \mu\mu$ and $J/\psi \rightarrow \mu\mu$ decays~\cite{PERF-2015-10}.
Jet candidates are reconstructed from three-dimensional energy clusters formed using both the
electromagnetic and hadronic calorimeters~\cite{PERF-2014-07}.
Clusters are grouped using the \antikt algorithm~\cite{Fastjet,Cacciari:2008gp} with a radius parameter $R = 0.4$.
The jet energy scale (JES) and resolution (JER) are first corrected to particle level
using MC simulation and then calibrated \insitu through \Zjet, \gjet, and multijet measurements~\cite{PERF-2016-04}.
 
Two levels of selection criteria are defined for leptons and jets; the looser ``baseline'' criteria and the tighter ``signal'' criteria.
Baseline objects are used for resolving ambiguities between overlapping objects, calculating the missing transverse momentum (\pTmiss) of an event,
and as inputs to the data-driven estimation of \fake-lepton events.
Baseline electrons are required to meet the ``loose and B-layer likelihood'' quality criteria~\cite{EGAM-2018-01}, satisfy $\pt > 10~\GeV$, and be within the ID acceptance ($|\eta| < 2.47$) but outside the barrel/endcap transition region of the electromagnetic calorimeter ($1.37 < |\eta| < 1.52$).
Baseline muons are required to meet the ``medium'' quality criteria~\cite{PERF-2015-10}, satisfy $\pt > 10~\GeV$, and fall within the MS acceptance ($|\eta|<2.7$).
Each baseline electron or muon is also required to have a trajectory consistent with the primary vertex to suppress pileup.
For this purpose, the transverse impact parameter ($d_0$) of a lepton is defined as the distance in the transverse plane between the beam-line and the closest point of the associated ID track.
The longitudinal impact parameter ($z_0$) then corresponds to the $z$-coordinate distance between that point and the primary vertex.
A selection of $|z_0 \sin\theta|<0.5~\text{mm}$, where $\theta$ is the polar angle of the track, is required for each lepton to ensure it is compatible with the primary vertex.
 
Baseline jets are required to satisfy $\pt > 20$~\GeV and fall within the full calorimeter acceptance ($|\eta| < 4.5$).
The identification of baseline jets containing $b$-hadrons ($b$-jets) is performed using the MV2 multivariate discriminant built using information from track impact parameters,
the presence of displaced secondary vertices, and the reconstructed flight paths of $b$- and $c$-hadrons inside the jet~\cite{FTAG-2018-01}.
The identification criteria are tuned to an average identification efficiency of 85\% as obtained for $b$-jets in simulated \ttbar events,
corresponding to rejection factors of 25, 2.7, and 6.1 for jets originating from light quarks and gluons, $c$-quarks, and $\tau$-leptons, respectively.
 
While photons are not used directly in the analysis, baseline photons are defined for use in the calculation of \pTmiss.
Baseline photons are required to meet the ``tight'' quality criteria~\cite{EGAM-2018-01}, satisfy $\pt > 25~\GeV$,
and fall within the ID acceptance ($|\eta| < 2.37$) and outside the calorimeter's transition region ($1.37 < |\eta| < 1.52$).
 
To aid in the correct reconstruction and identification of leptons and jets an overlap-removal procedure is performed, preventing the reconstruction of a single particle as multiple objects.
First, any electron that shares a track with a muon in the ID is removed, as the track is consistent with track segments in the MS.
Next, jets are removed if they are within $\Delta R = 0.2$ of a lepton and are either not $b$-tagged or satisfy $\pt > 100~\GeV$,
as they are consistent with the energy deposited by an electron shower or muon bremsstrahlung.
For the overlap of a jet with a nearby muon, the jet is discarded only if it is associated with fewer than three tracks of $\pt \geq 500~\MeV$.
Finally, electrons and muons within $\Delta R = 0.4$ of any remaining jets are discarded to reject \fake leptons originating from hadron decays.
In the overlap-removal procedure the calculation of $\Delta R$ uses rapidity instead of $\eta$ to ensure the distance measurement is Lorentz invariant for jets with non-negligible masses.
 
The \pTmiss of each event, with magnitude \met, is defined as the negative vector sum of the
transverse momenta of all identified baseline objects (electrons, muons, jets, and photons) and an additional soft term~\cite{PERF-2016-07}.
The soft term is constructed from all tracks associated with the primary vertex that are not associated with any baseline object.
The \pTmiss is therefore adjusted to include the full calibration of the reconstructed baseline objects while minimizing any pileup dependence in the soft term.
 
Tighter ``signal'' criteria are applied to the final leptons and jets considered by the analysis to ensure a high selection purity and accurate \pt measurement.
Any event with a baseline lepton that fails to satisfy the signal criteria is rejected to reduce the contamination from \fake-lepton events.
Signal leptons are required to have $\pt>12~\GeV$ and electrons must meet the ``medium'' quality criteria~\cite{EGAM-2018-01}.
At least one of the signal leptons must be identified as having activated a trigger and must pass the larger \pt requirement of that trigger.
The track associated with each signal electron or muon must pass a requirement on $d_0$ and its uncertainty $\sigma_{d_0}$
such that $|d_0/\sigma_{d_0}| < 5~(3)$ for electrons (muons), ensuring the selection of leptons with prompt, well-reconstructed tracks.
Finally, signal leptons must be sufficiently isolated from additional detector activity by passing a \pt-dependent ``tight'' requirement on both calorimeter-based and track-based isolation variables~\cite{EGAM-2018-01,PERF-2015-10}.
The calorimeter-based isolation is defined within a cone of size $\Delta R = 0.2$ around the lepton,
and the amount of nonassociated calorimeter transverse energy within the cone must be below 6\% (15\%) of the electron (muon) \pt.
The track-based isolation cone size is $\Delta R = 0.2$ for low-\pt electrons and decreases linearly with \pt above 50~\GeV as the electron's shower becomes more collimated.
For muons, the size of the track-isolation cone is $\Delta R = 0.3$ for muons with $\pT \leq 33~\GeV$ and decreases linearly with \pt to $\Delta R = 0.2$ at \pt = 50~\GeV,
improving the selection efficiency for higher-\pt muons.
The track-based isolation only considers nonassociated tracks that are consistent with the primary vertex, and the scalar sum of track \pt (\ptiso) is required to be below 6\% (4\%) of the electron (muon) \pt.
The lepton \ptcone is then defined as the scalar sum of the lepton \pt and \ptiso, and is useful in parameterizing the behavior of \fake leptons.
The muon ``tight'' isolation requirement is roughly 96\% efficient for all \pt and $\eta$~\cite{PERF-2015-10}, while for electrons it is 70\% efficient at 20~\GeV and becomes more than 98\% efficient above 100~\GeV~\cite{EGAM-2018-01}.
 
Signal jets are required to have $|\eta| < 2.8$, and events are rejected if they contain a jet that fails to meet the ``loose'' quality criteria~\cite{ATLAS-CONF-2015-029},
reducing contamination from electronic noise bursts and noncollision backgrounds.
To suppress jets originating from pileup, jets with $\pT < 120$~\GeV and within the ID acceptance ($|\eta| < 2.5$) are required to pass the
``medium'' working point of the track-based jet vertex tagger~\cite{PERF-2014-03,ATLAS-CONF-2014-018}.
All MC simulation samples are corrected using per-event weights to account for small differences with data in signal-lepton identification, reconstruction, isolation and triggering efficiencies~\cite{EGAM-2018-01,PERF-2015-10},
as well as in signal-jet pileup rejection~\cite{PERF-2014-03} and flavor-identification efficiencies~\cite{FTAG-2018-01}.
 
\section{Search strategy}
\label{sec:search}
 
The \BL RPV model allows for many different decay modes of \chono and therefore many possible final states.
A decay of interest is \triLepDecay because of the large number of leptons produced from a single resonance.
The invariant-mass distribution of the trilepton resonance (\mZl) is narrow due to the excellent momentum resolution of reconstructed electrons and muons.
No SM process naturally produces a three-lepton resonance, leading to a smooth combinatorial background distribution in which a resonance would be distinguishable.
 
Three orthogonal signal regions (SRs) are developed in Section~\ref{sec:search:SR} to select \CCsignal and \CNsignal events with at least one \triLepDecay decay.
Each SR targets different decay scenarios of the second \chono through requirements on the number of leptons and reconstructed $W$, $Z$, or Higgs bosons.
Matching procedures are developed for events with additional leptons or boson candidates to optimally assign the decay products to each \chono, as described in Section~\ref{sec:search:DirectMatching}.
The SRs utilize event-wide information to reduce combinatorial backgrounds, as described in Section~\ref{sec:search:BkgRej}.
 
\subsection{Signal regions targeting trilepton decays}
\label{sec:search:SR}
 
Each SR requires at least three signal leptons, two of which are identified as candidate $Z$ boson decay products if they have
the same flavor and opposite sign of their electric charge (SFOS) and have an invariant mass \mll within 10~\GeV of the $Z$ boson mass.
For events which have more than one SFOS pair, the pair with \mll closest to the $Z$ mass is chosen.
The \mZl of the \chone is then reconstructed from the chosen SFOS pair and a third lepton.
Deviations of \mll from the expected $Z$ boson mass of 91.2~\GeV can occur due to the imperfect energy reconstruction of leptons, particularly at high \pt.
The \mZl resolution is therefore improved by shifting the value of \mZl by an amount equal to ($91.2-\mll$)~\GeV.
 
\begin{figure}[htb!]
\centering
\includegraphics[width=0.6\textwidth]{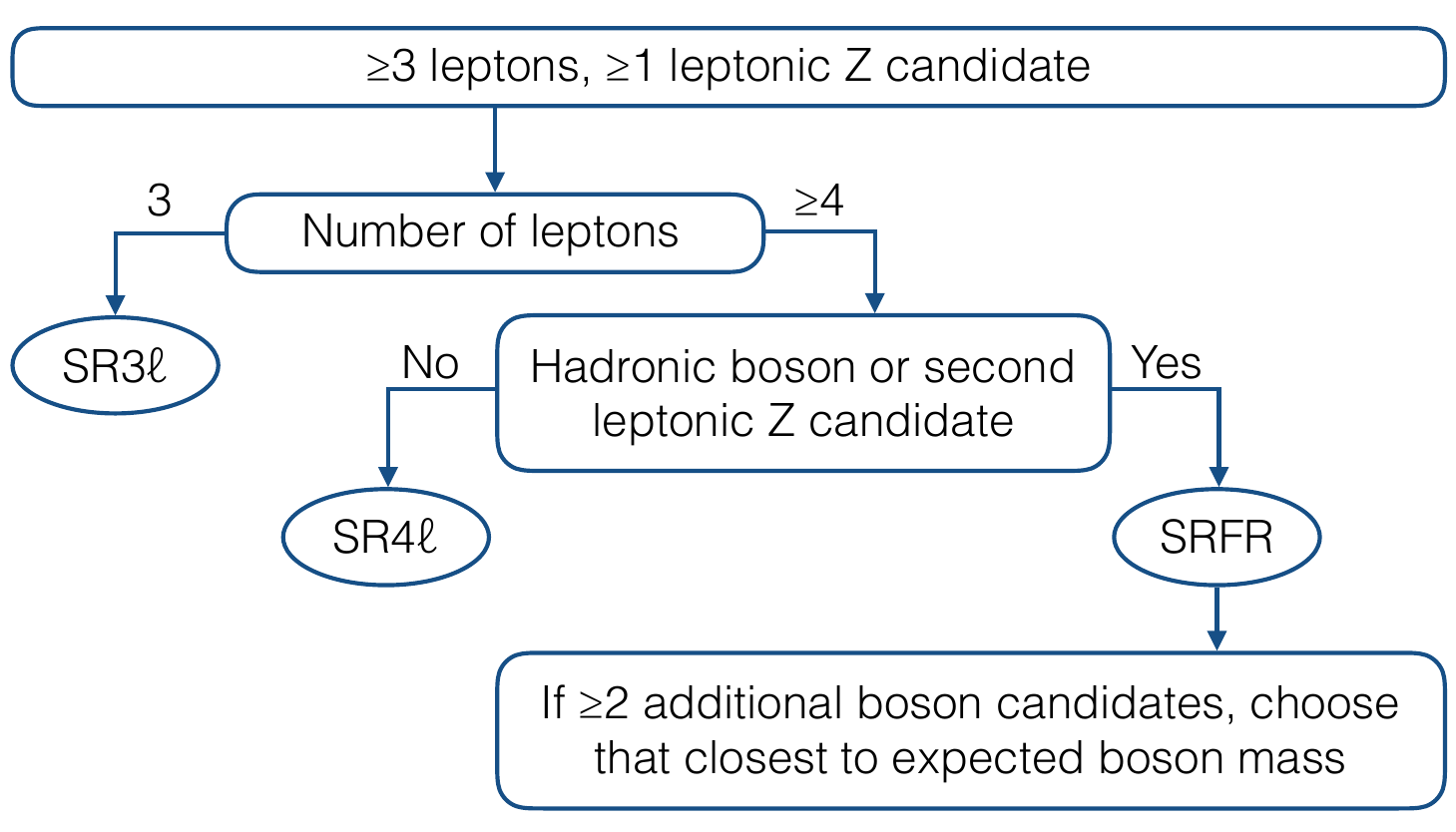}
\caption{Schematic flow chart describing the assignment of an event to a given signal region.}
\label{fig:search:SRPlacement}
\end{figure}
 
Events are separated into the three SRs according to the number of leptons and the presence of a second reconstructed $Z$, $W$, or Higgs boson from the second \chono decay.
The \SRTL region targets events where all decay products are visible and ``fully reconstructed''.
The \SRFour region targets events with four or more leptons and possible \MET,
while the \SRThree region targets events with only three visible leptons and substantial \MET,
with at least one neutrino coming from the decay of the second \chono.
The choice of SR for an event is described below and summarized in Figure~\ref{fig:search:SRPlacement}.
Additional selections to reduce the SM background contributions are subsequently applied in each of the SRs separately, as described in Section~\ref{sec:search:BkgRej}.
 
To target fully visible events, \SRTL requires a fourth lepton and a second reconstructed $Z$, $W$, or Higgs boson.
Pairs of jets are considered for the second boson if their invariant mass \mjj is consistent with that of a $W$ or $Z$ boson, with $71.2 < \mjj < 111.2~\GeV$.
If at least one of the jets is a $b$-jet, the \mjj requirement is loosened to $71.2 < \mjj < 150~\GeV$ to allow for Higgs boson decays.
Additional SFOS lepton pairs are also considered for the second boson candidate in events with six or more leptons if their invariant mass is consistent with the $Z$ boson mass, such that $81.2 < \mll < 101.2~\GeV$.
If there are multiple candidates for the second boson, the pairing selected is that
with invariant mass closest to the $Z$ boson mass, or closest to the Higgs boson mass for pairs that include at least one $b$-jet.
 
The \SRFour region targets events in which the decay of the second \chono includes one or more leptons but is not fully reconstructed due to the presence of neutrinos.
Events with four or more leptons that fail all \SRTL requirements are selected by \SRFour.
The \SRThree region targets decays of the second \chono that include no leptons, requiring exactly three leptons in the event.
While each region targets specific \chono decay chains, events in which one or more leptons fall outside the detector acceptance or are not reconstructed may still be selected by other regions.
For the signal sample with a mass of 500~\GeV and democratic \chono branching fractions to bosons and leptons, the \SRTL, \SRFour, and \SRThree regions have selection efficiencies of 3\%, 4\%, and 5\%, respectively.
 
Within each SR the search is performed in the \mZl spectrum to maximize the discovery sensitivity to a resonance.
The binning of the \mZl observable was optimized using simulated \CCsignal and \CNsignal signal samples with reconstructed invariant mass resolutions of around 2\%,
as measured from the widths of Gaussian fits to the reconstructed invariant mass distributions.
The optimized binning accounts for the predicted background expectation.
Lower edges are set at
\begin{equation}
\mZl = 90, 110, 130, 150, 170, 190, 210, 230, 250, 270, 300, 330, 360, 400, 440, \mathrm{~and~}580~\GeV.
\label{eq:binning}
\end{equation}
The last bin has no upper edge and includes all events with $\mZl > 580~\GeV$.
The same binning is used for all three SRs, facilitating the discovery of a trilepton resonance that would contribute to all SRs.
 
\subsection{Assignment of leptons and boson candidates to \chono decays}
\label{sec:search:DirectMatching}
 
The presence of one or more additional leptons from the second \chono decay introduces ambiguity in the assignment
of a lepton and boson produced directly from a \chono decay.
A matching procedure is implemented to identify the ``direct'' leptons that come directly from the \chono decays, rather than from the subsequent decay of a boson, and to assign them to each \chono.
The procedure optimizes the sensitivity to signals of various masses by maintaining a high efficiency for the correct assignments while reducing the contamination from SM processes.
In \SRTL, both the trilepton decay and the fully visible decay of the second \chono, with reconstructed mass \mBl, are chosen as the groupings
that minimize the mass asymmetry between the mass-degenerate \CCsignal or \CNsignal pair, where \mZlAsym is defined as
\begin{equation}
\mZlAsym = \frac{|\mZl - \mBl|}{\mZl + \mBl}.
\label{eq:asym}
\end{equation}
The matching efficiency for the signal samples is 60\% at 100~\GeV and 80\% or more for masses of 200~\GeV and larger.
 
The matching procedure for a direct lepton to the $Z$ candidate for all other analysis regions with four or more leptons is developed to optimize the sensitivity of the \SRFour region.
Two methods are implemented, and the choice of method exploits the correlation between the true mass of the \chono and \Lt, the scalar sum of the \pt of all leptons in the event.
A method targeting low-mass signals is used when $\Lt<550~\GeV$ and a method targeting high-mass signals is used when $\Lt\ge550~\GeV$.
For low-mass signals, the \chono can often be produced with a sufficiently large momentum such that the decay products are near to one another,
and the lepton that is closest in angular distance $\DeltaR$ to the reconstructed $Z$ boson is chosen.
For high-mass signals, the \chono decay products are often produced at a wide angle with respect to each other,
and mispairings will produce a \mZl that is smaller than the \chono mass.
Therefore, the lepton that maximizes the reconstructed \mZl is chosen.
The matching efficiency of this procedure for signal samples with various \chono masses is 90\% at 100~\GeV, 30\% at 300~\GeV, and 70\% at 700~\GeV.
While a low matching efficiency is seen at 300~\GeV due to the use of $\DeltaR$ matching when the \mZl maximization would be preferred,
the overall analysis sensitivity is improved by avoiding the \mZl maximization of low-\Lt backgrounds.
 
As noted in Section~\ref{sec:intro}, the preferred flavor of the direct lepton(s) is related to the neutrino mass hierarchy.
The sensitivity to \CCsignal and \CNsignal events may therefore be improved by imposing constraints on the flavor of the direct lepton(s),
targeting the favored signal decays while rejecting additional SM backgrounds.
Two additional sets of SRs are developed that are each identical to the nominal set of three SRs
except that they require the direct lepton(s) to be either electron (\SRTLE, \SRFourE, \SRThreeE) or muon (\SRTLM, \SRFourM, \SRThreeM).
These additional ``$e$'' and ``$\mu$'' channels are used separately from the ``inclusive'' channel and from one another,
and are only used when targeting signal models with high \chono branching fractions to either electrons or muons, as discussed in Section~\ref{sec:search:exclusion}.

\subsection{Rejection of combinatorial Standard Model backgrounds}
\label{sec:search:BkgRej}
 
The composition and kinematics of the final-state particles that are produced from the decay chains of the \CCsignal or \CNsignal processes can be combinatorially reproduced by certain SM processes.
The \ZZ process has a significant contribution to \SRTL and \SRFour when both $Z$ bosons decay leptonically.
Events from the \ZZ process are rejected if they have exactly four leptons that form two SFOS pairs and the mass \mllTwo of the second pair, the pair not selected for the primary \chone candidate, is within 20~\GeV of the $Z$ boson mass.
In \SRFour, which targets decay chains of the second \chono with at least one neutrino, the \ZZ contribution is further reduced by requiring $\met>80~\GeV$ in events with a second same-flavor lepton pair.
 
The SM \ttZ process can also contribute significantly in the SRs, and is identifiable by the presence of two $b$-jets
from the two top-quark decays.
Signal events that include a Higgs boson decay may also include two $b$-jets, with a 72\% efficiency of identifying both $b$-jets using the flavor-tagging algorithm described in Section~\ref{sec:eventreco}.
The $b$-jets will often be collimated due to the boost of the Higgs boson.
Therefore, an additional selection is applied in all SRs that requires the two highest-\pt $b$-jets, if they are found in the event, to satisfy $\dRbb<1.5$.
 
The \ZZ, \ttZ, and other SM backgrounds can be further reduced in \SRTL by taking advantage of the fully visible decay of the second \chono.
As the \chone and \none are expected to be mass-degenerate, the \mZlAsym (Eq.~(\ref{eq:asym})) between the \CCsignal or \CNsignal pair is expected to be small.
A requirement of $\mZlAsym < 0.1$ in \SRTL is effective in rejecting combinatorial backgrounds for which \mZlAsym is more evenly distributed.
 
Events in \SRThree are expected to exhibit a significant \met because the second \chono decays directly into a neutrino and a boson, while the subsequent decay of the boson may also produce neutrinos.
A requirement of $\met>150~\GeV$ reduces contamination from SM processes with no neutrinos, particularly \Zjet events that include a \fake lepton.
The SM \WZ process with fully leptonic decays is also a significant contributor to \SRThree, and contains a single neutrino from the $W$ decay.
The measured \met is therefore representative of the \pt of the neutrino, and the transverse mass \mT of the $W$ boson can be reconstructed from the \pt of the lepton and the
azimuthal separation $\Delta\phi$ between the lepton and \pTmiss, with
\begin{equation*}
\mT = \sqrt{2\pt\met(1-\cos(\Delta\phi))}.
\end{equation*}
The \mT of a $W$ boson has a kinematic edge at the $W$ mass, and signal events in \SRThree usually produce lepton--\met pairings with a larger \mT.
The minimum \mT of all lepton--\met pairings for which the other two leptons form a SFOS pair, defined as \mTmin, is required to be $\mTmin>125~\GeV$ in \SRThree.
This definition allows \WZ events to be rejected even if the incorrect SFOS pair was selected for the $Z$ boson.
 
\section{Background estimation and validation}
\label{sec:background}
The MC samples described in Section~\ref{sec:datamc} are used to predict the expected background yield from SM processes.
To improve the accuracy of the MC prediction in the unique phase-space of this analysis and to constrain the systematic uncertainties discussed in \secref{systematics},
the MC predictions are normalized in control regions (CRs).
Each CR is dedicated to the measurement of an important SM process and they are discussed in \secref{background:MC}.
A dedicated data-driven estimation is used for the \fake-lepton background and is discussed in \secref{background:Fake}.
A fit based on a profile likelihood test statistic~\cite{Cowan:2010js} is performed on all CRs and SRs simultaneously using the HistFitter package~\cite{Baak:2014wma}
to estimate the final post-fit background prediction and uncertainty.
 
\begin{table}[ht!]
\small
\begin{center}
\caption{Selection criteria for the various signal, control, and validation regions used in the analysis.
All regions require a pair of leptons with the same flavor and opposite sign of their electric charge and with an invariant mass between 81.2~\GeV and 101.2~\GeV.
Additionally, they require a third lepton and a trilepton invariant mass above 90~\GeV.
The 2nd boson requirement indicates the presence of two additional jets or leptons consistent with a $W$, $Z$, or Higgs boson decay.
The asterisk (*) in the \SRFour \Met requirement indicates that this selection is only considered for events with two pairs of same-flavor leptons.
The \dRbb selection is only considered for events with at least two $b$-jets.
}
\label{tab:regionDefs}
\begin{tabular}{c|cccccccc}
\toprule
\multirow{2}{*}{Region} & \multirow{2}{*}{$N_{\mathrm{lep}}$} & \multirow{2}{*}{\Met [GeV]} & \multirow{2}{*}{\mTmin [GeV]} & 2nd & 2nd leptonic $Z$; & \multirow{2}{*}{$N_{b\textrm{-jet}}$} & \multirow{2}{*}{$\dRbb$} & \multirow{2}{*}{\mZlasym} \\
& & & & boson & $|\mllTwo-m_{Z}|$ [GeV] & & & \\
\midrule
\textbf{\SRTL}     & $\geq$4 & -       & -         & Yes & veto; <20       & -       & <1.5      & < 0.1 \\
\textbf{\SRFour}   & $\geq$4 & >80*    & -         & No  & veto; <20       & -       & <1.5      & -     \\
\CRZZ     & =4     & -       & -         & -   & require; <5     & -       & <1.5      & -     \\
\VRZZ     & =4     & -       & -         & -   & require; [5,20] & -       & <1.5      & -     \\
\CRttZ    & $\geq$3 & >40     & -         & -   & veto; <20       & $\geq$2 & >2.5      & -     \\
\VRttZ    & $\geq$3 & >40     & -         & -   & veto; <20       & $\geq$2 & [1.5,2.5] & -     \\
\midrule
\textbf{\SRThree}  & =3     & >150    & >125      & -   & -               & -       & <1.5      & -     \\
\CRWZ     & =3     & <80     & [50,100]  & -   & -               & -       & <1.5      & -     \\
\VRmet    & =3     & >80     & <100      & -   & -               & -       & <1.5      & -     \\
\VRmTmin  & =3     & <80     & >125      & -   & -               & -       & <1.5      & -     \\
\midrule
\CRZj     & =3     & <30     & <30       & -   & -               & -       & <1.5      & -     \\
\VRZj     & =3     & [30,80] & <30       & -   & -               & -       & <1.5      & -     \\
\bottomrule
\end{tabular}
\end{center}
\end{table}
 
The CRs are developed to be kinematically similar to the SRs but with a small number of selections inverted,
reducing any possible signal contamination and ensuring orthogonality between regions.
Validation regions (VRs) between the CRs and SRs are developed to ensure the validity of the extrapolation of the yield normalization across the inverted selections and into the SRs.
The regions are developed so that any possible signal contamination from \chono with democratic branching fractions to bosons and leptons
is typically less than 1\% in each CR and less than 5\% in each VR.
This ensures an accurate estimation of the SM backgrounds and an unbiased validation.
Any contamination from the signal model in the CRs is accounted for in the fit for completeness.
All regions are required to have at least three leptons and one SFOS pair with \mll within 10~\GeV of the $Z$ boson mass.
The CRs and VRs are inclusive in \mZl as this variable is seen to be well modeled by the MC simulation.
A requirement of $\mZl > 90~\GeV$ is made in all regions, corresponding to the lowest \mZl probed by the SRs.
The selections for the various regions are discussed below and summarized in Table~\ref{tab:regionDefs}.

\subsection{Primary backgrounds}
\label{sec:background:MC}
 
The major SM backgrounds that are fitted in dedicated CRs are the \WZ, \ZZ, and \ttZ processes.
The yields of other SM processes are small and are therefore not normalized by the fit but taken directly from the MC prediction.
These include the triboson, Higgs boson, and ``Other'' background categories, where Other consists almost completely of the \tWZ, \ttW, and \tZ processes.
 
The \WZ process is dominant in the three-lepton \SRThree, and the \CRWZ control region is developed by inverting the \MET requirement and selecting events with \mTmin consistent with the presence of a $W$ boson.
This removes possible signal contamination from \chone and \none, which typically have a high \MET and \mTmin in \SRThree due to one or more boosted neutrinos.
Two VRs, \VRmet and \VRmTmin, are designed to test the validity of the \WZ normalization in \SRThree using similar \met and \mTmin requirements, respectively.
Good data--MC agreement is seen in both VRs, and the \met and \mTmin distributions are shown for \CRWZ, \VRmet, and \VRmTmin in Figure~\ref{fig:CRVR}.
These distributions have all region selections applied except the variable shown, where the CR or VR selections are indicated by arrows.
The exception is the \MET distribution in \VRmet, which is shown with all region selections applied.
The underflow of the \mTmin distribution does not consider events with \mTmin below 30~\GeV.
 
\begin{figure}[hb!]
\centering
\includegraphics[width=0.45\textwidth]{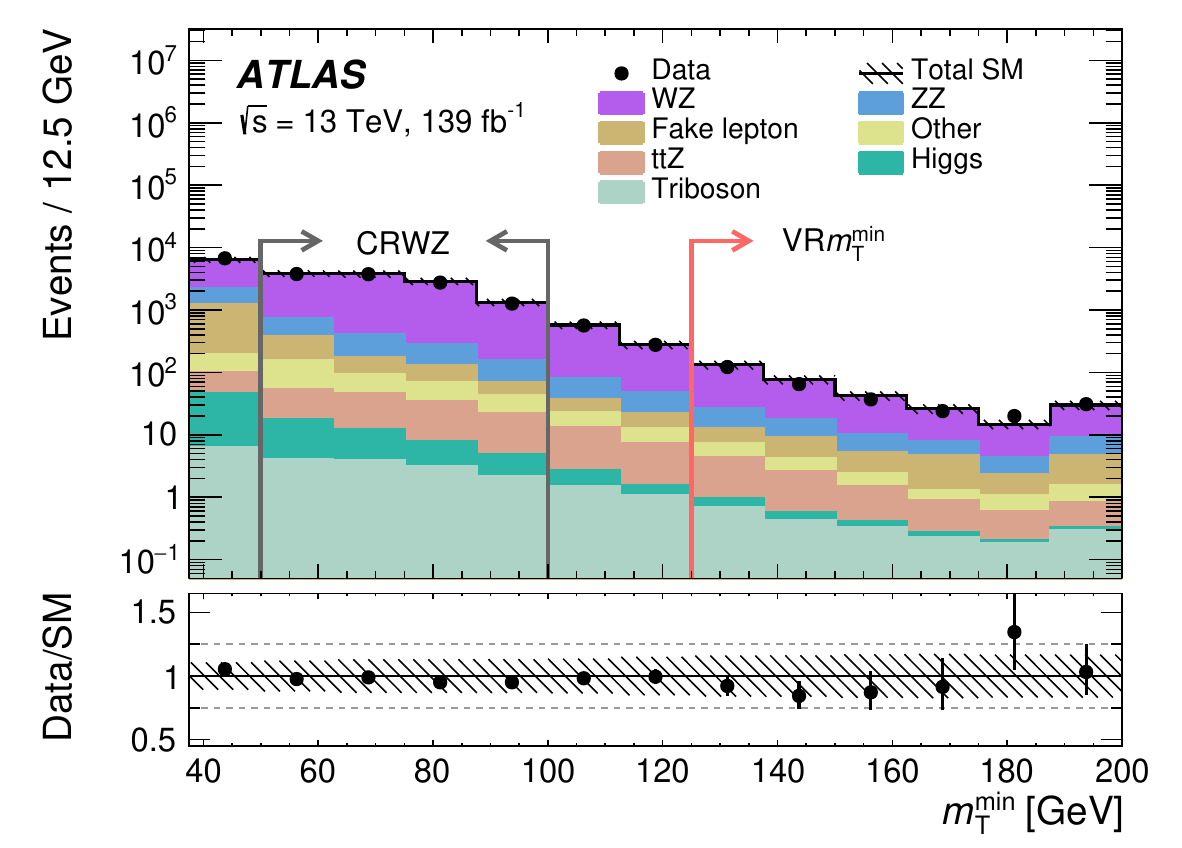}
\includegraphics[width=0.45\textwidth]{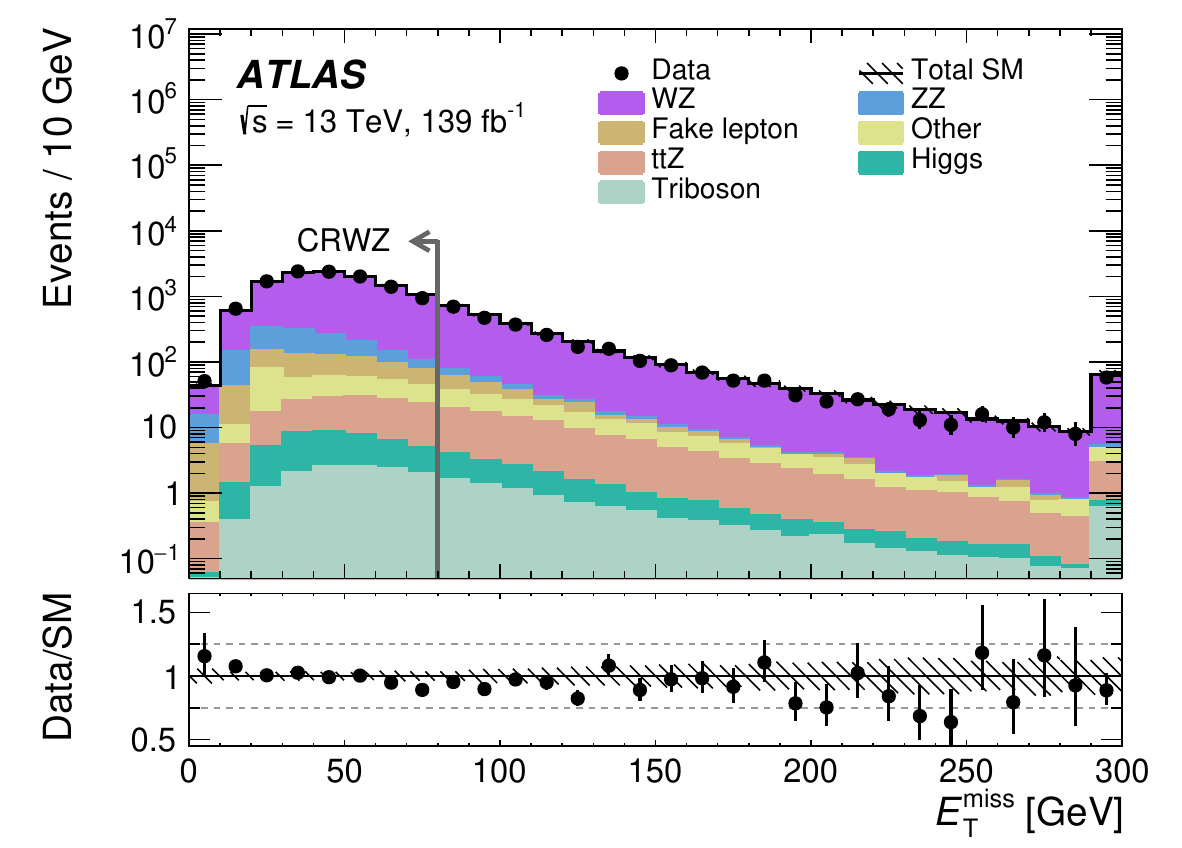}
\includegraphics[width=0.45\textwidth]{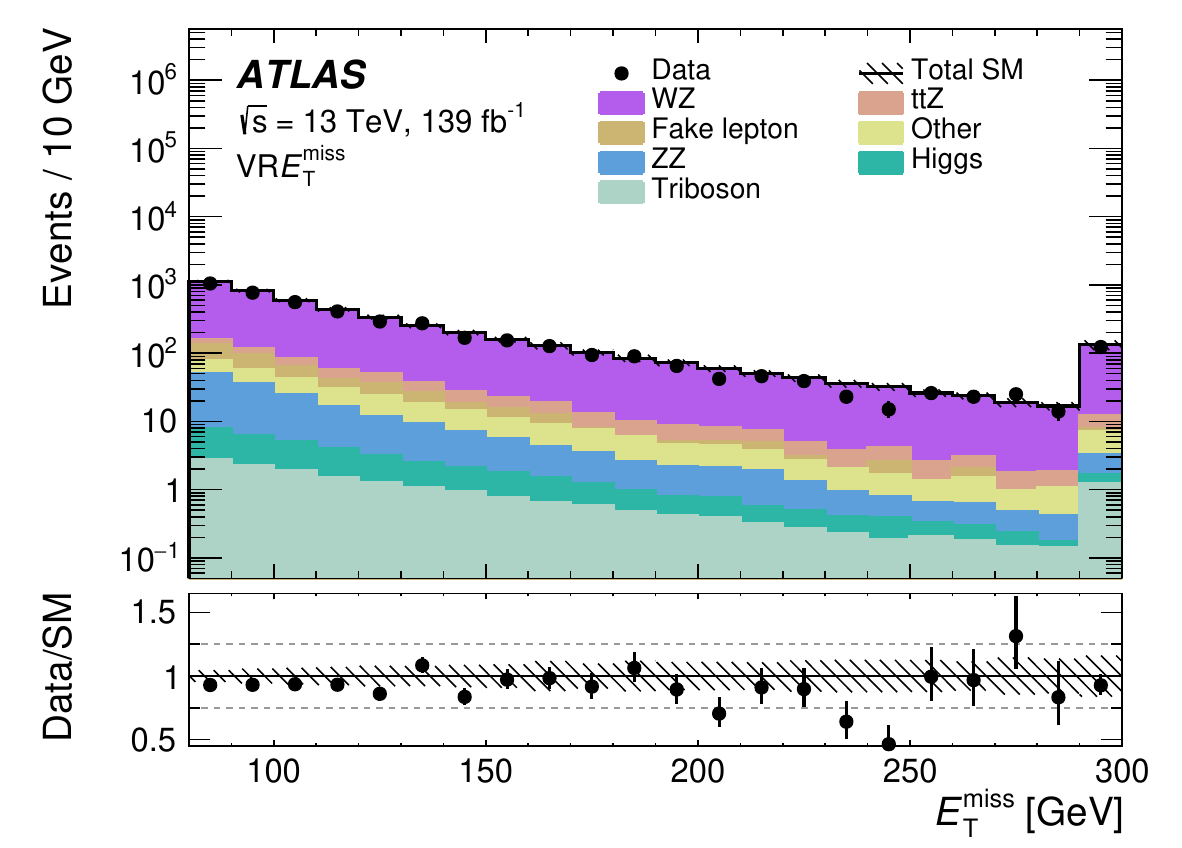}
\includegraphics[width=0.45\textwidth]{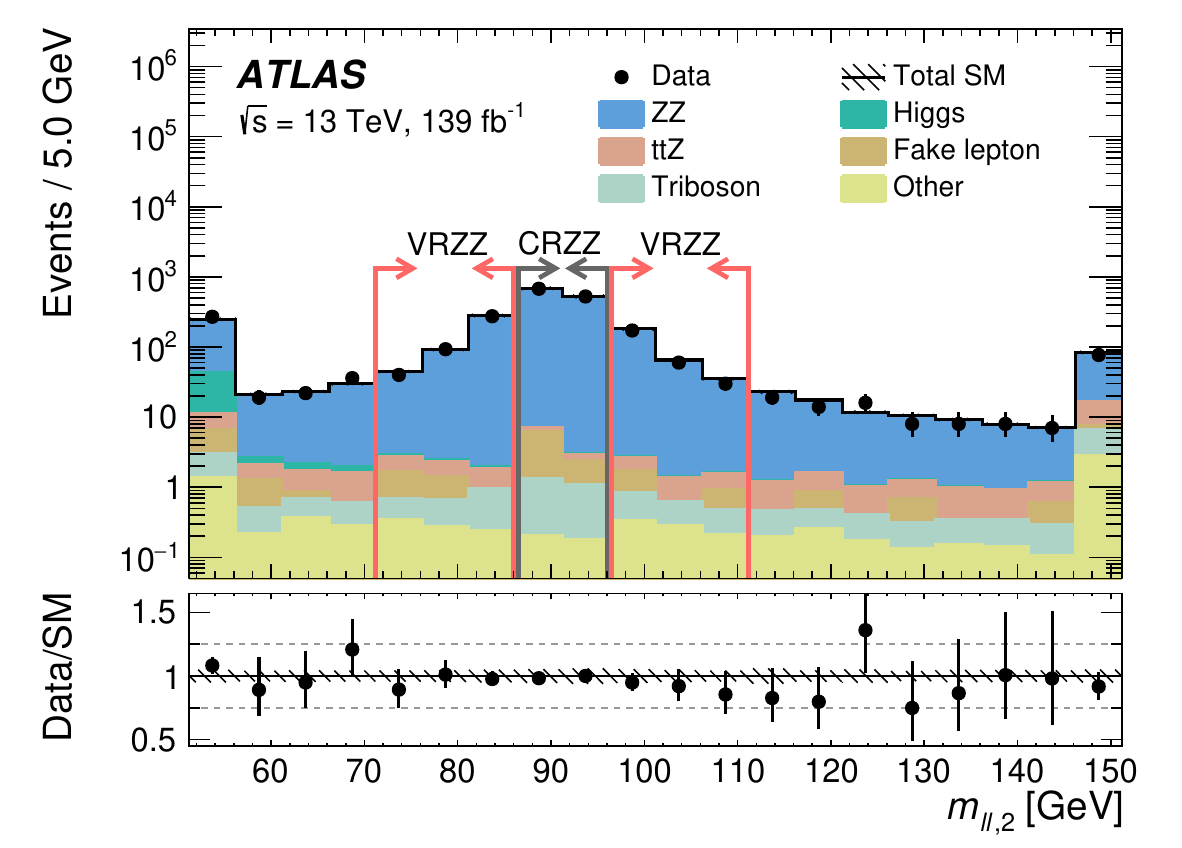}
\includegraphics[width=0.45\textwidth]{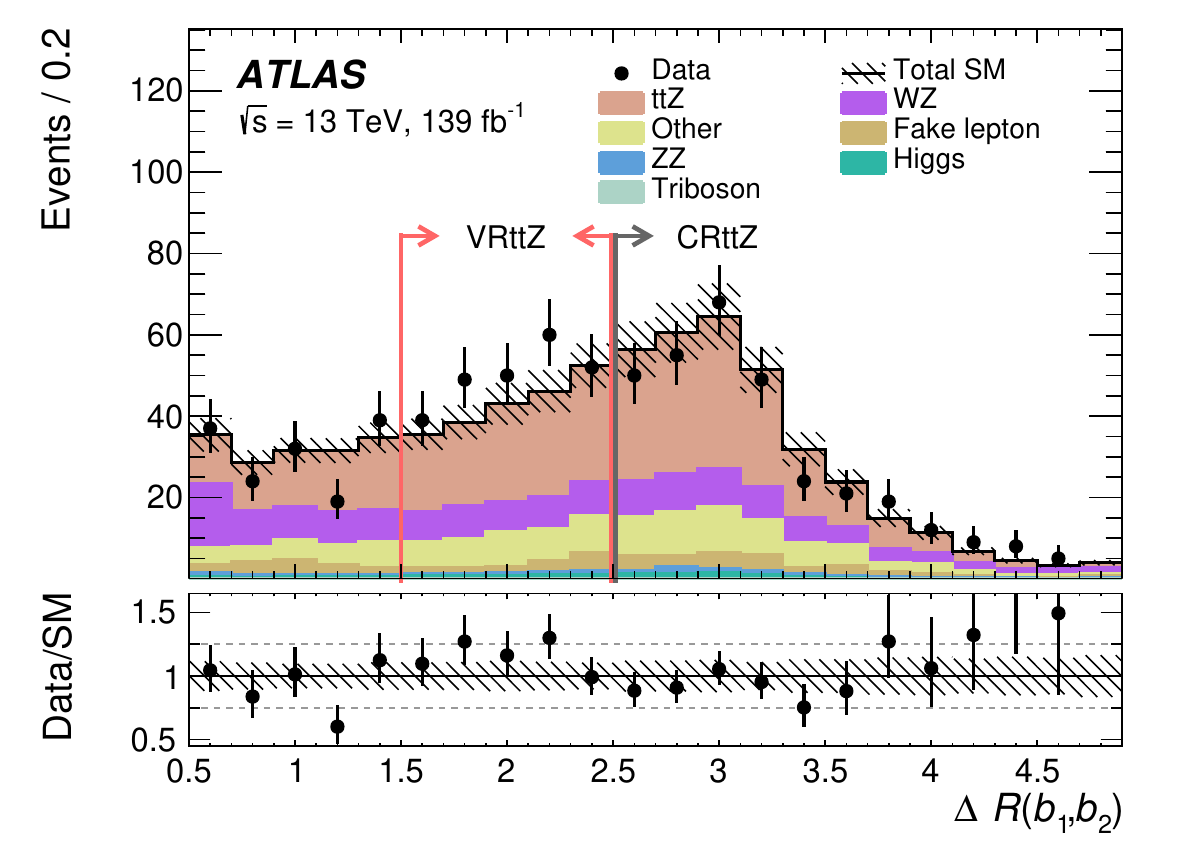}
 
\caption{
Distributions of the data and post-fit background in the CRs and VRs that are relevant in the extrapolation to the SRs, including
(top left) \mTmin in \CRWZ and \VRmTmin, (top right) \MET in \CRWZ, (middle left) \MET in \VRmet,
(middle right) \mllTwo in \CRZZ and \VRZZ, and (bottom) \dRbb in \CRttZ and \VRttZ.
Black (red) arrows indicate the CR (VR) selection on the variable shown, with all other region selections applied.
\captionoverflow
\captionlegend
\captionsysband
\captionratio
}
\label{fig:CRVR}
\end{figure}
 
\begin{figure}[hb!]
\centering
\includegraphics[width=0.399\textwidth]{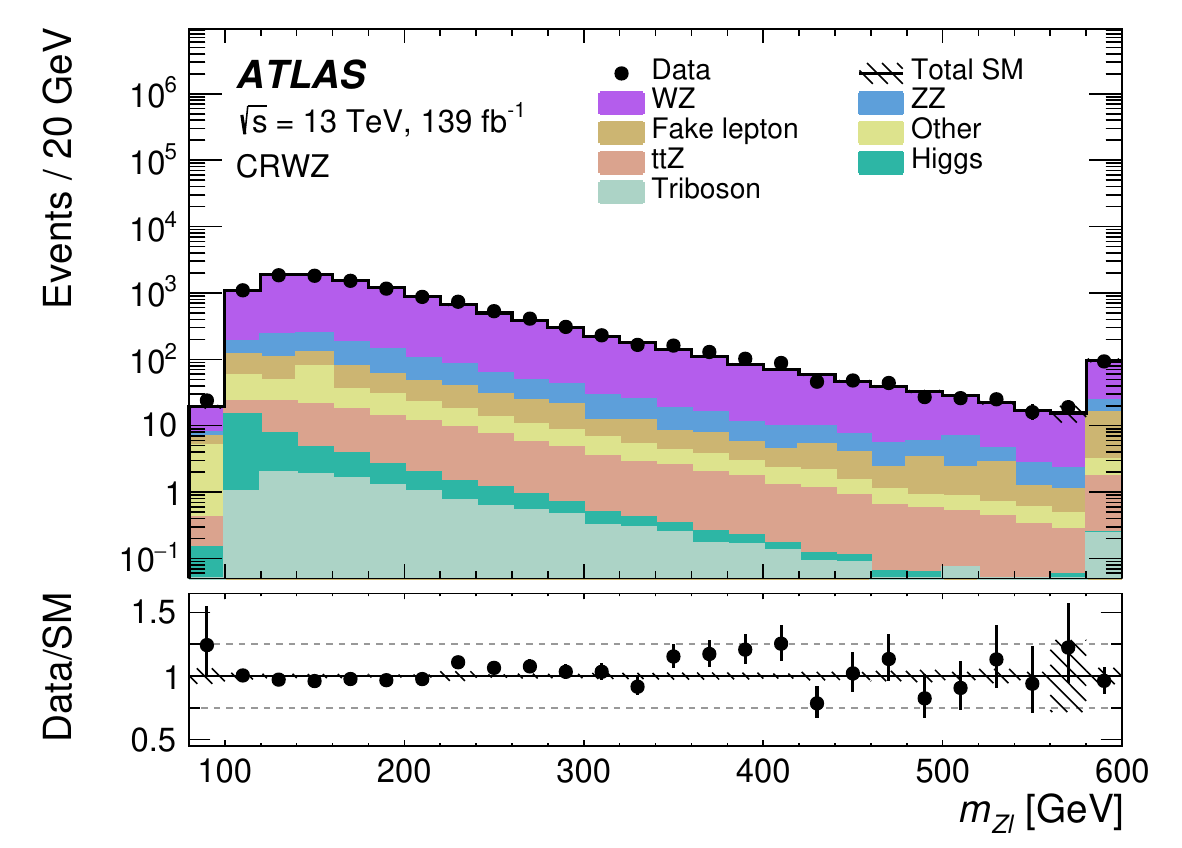}
\includegraphics[width=0.399\textwidth]{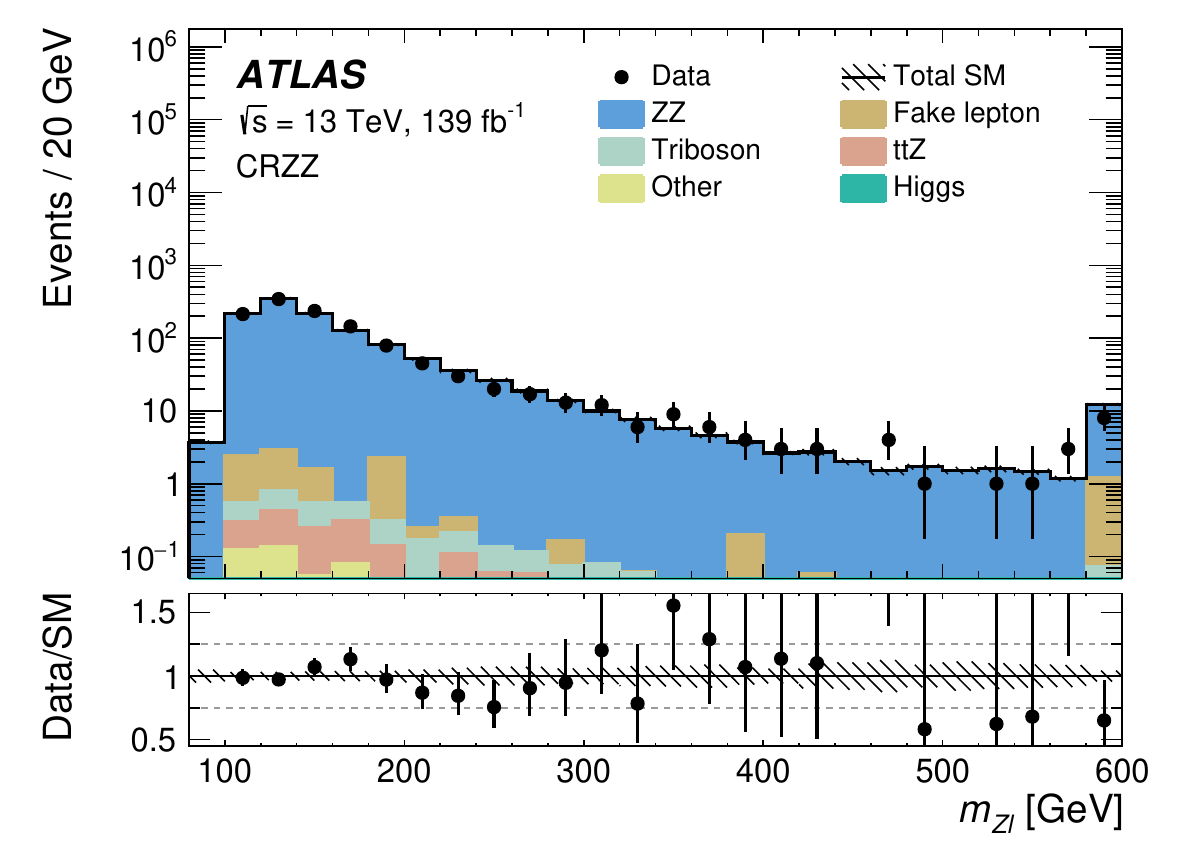}
\includegraphics[width=0.399\textwidth]{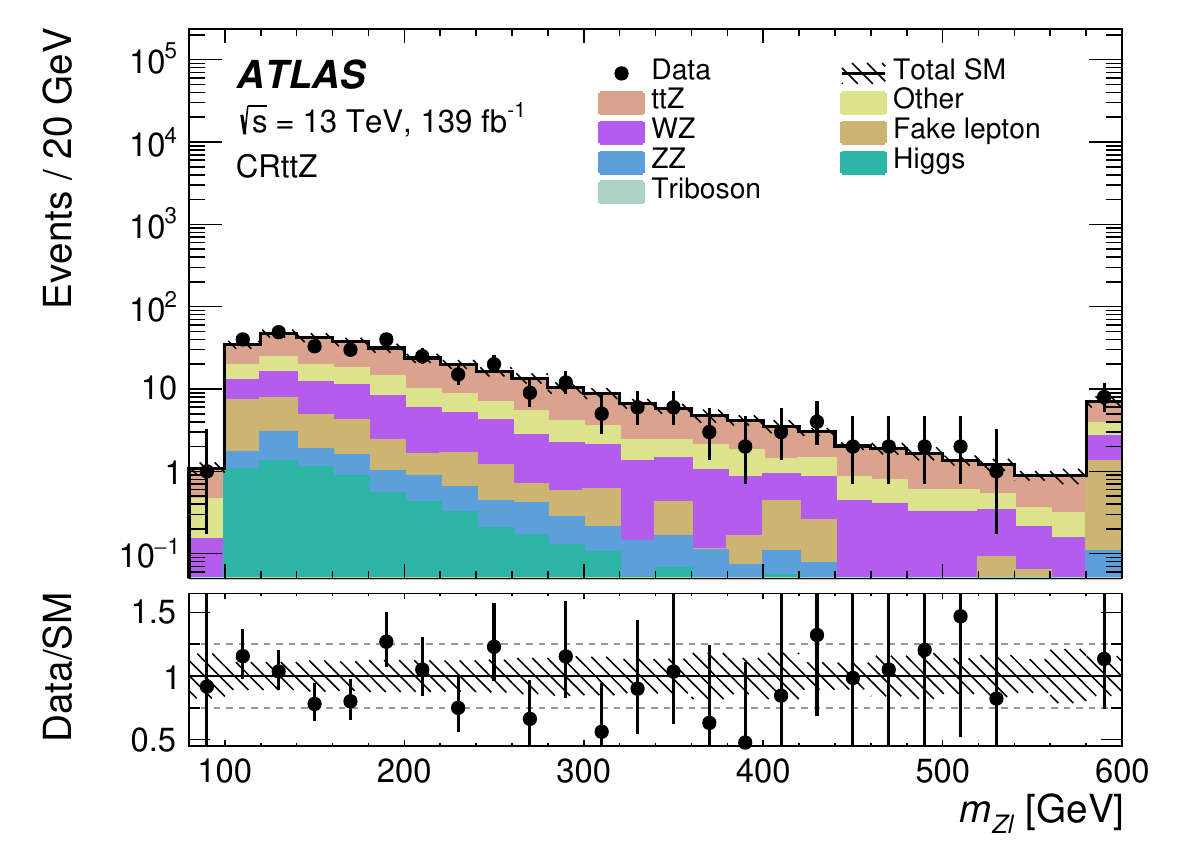}
\includegraphics[width=0.399\textwidth]{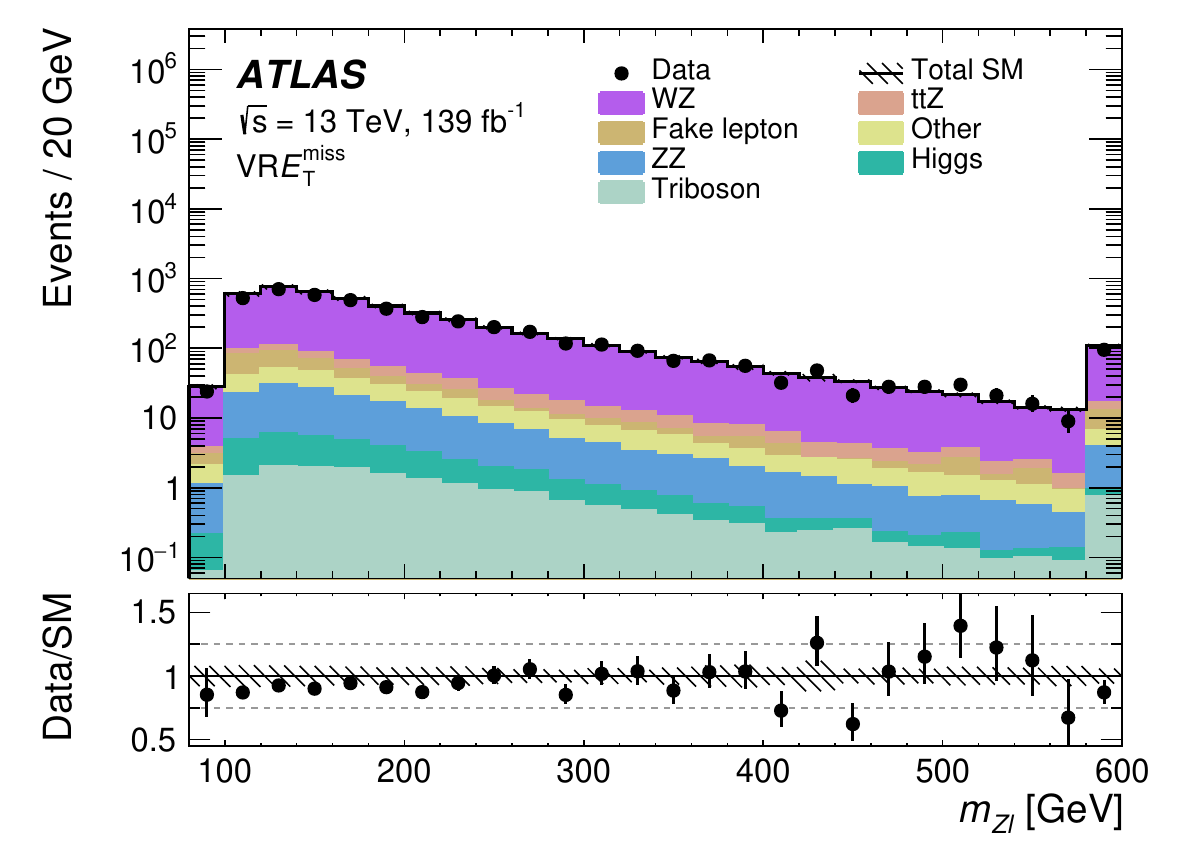}
\includegraphics[width=0.399\textwidth]{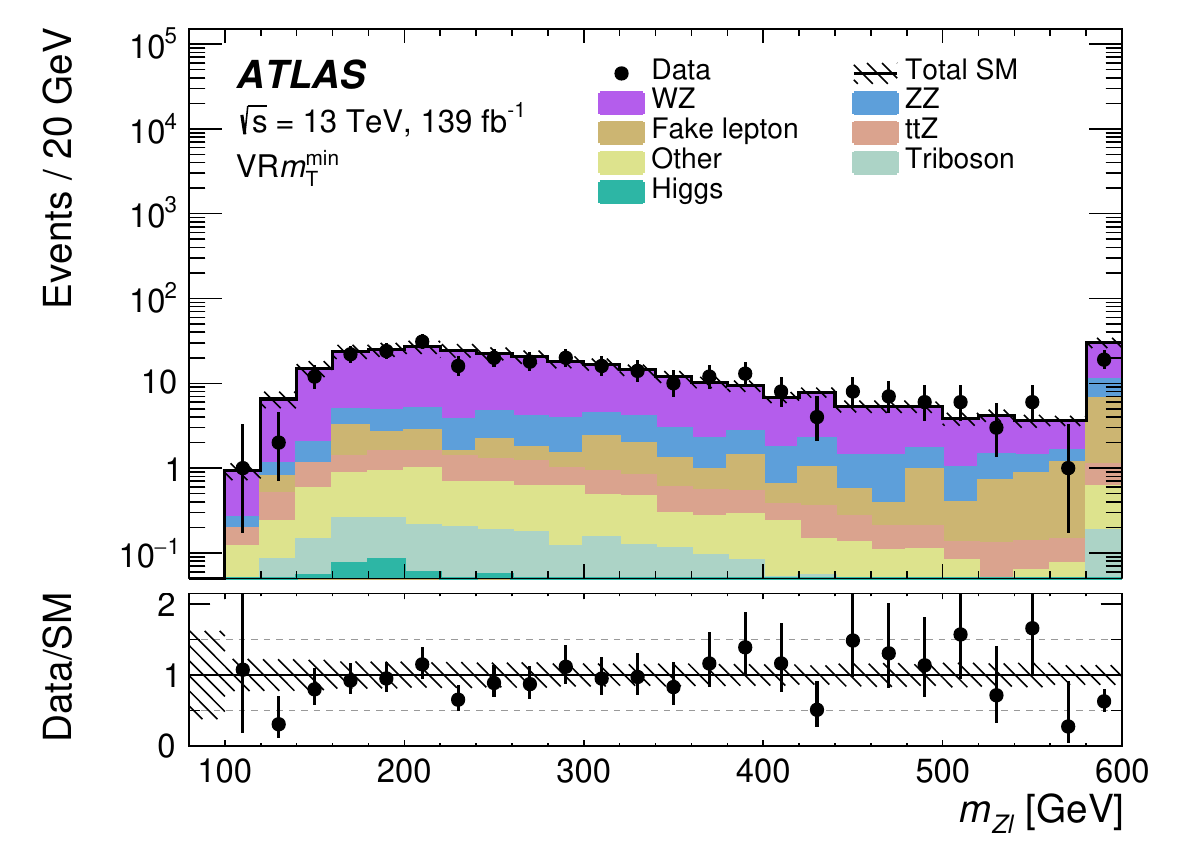}
\includegraphics[width=0.399\textwidth]{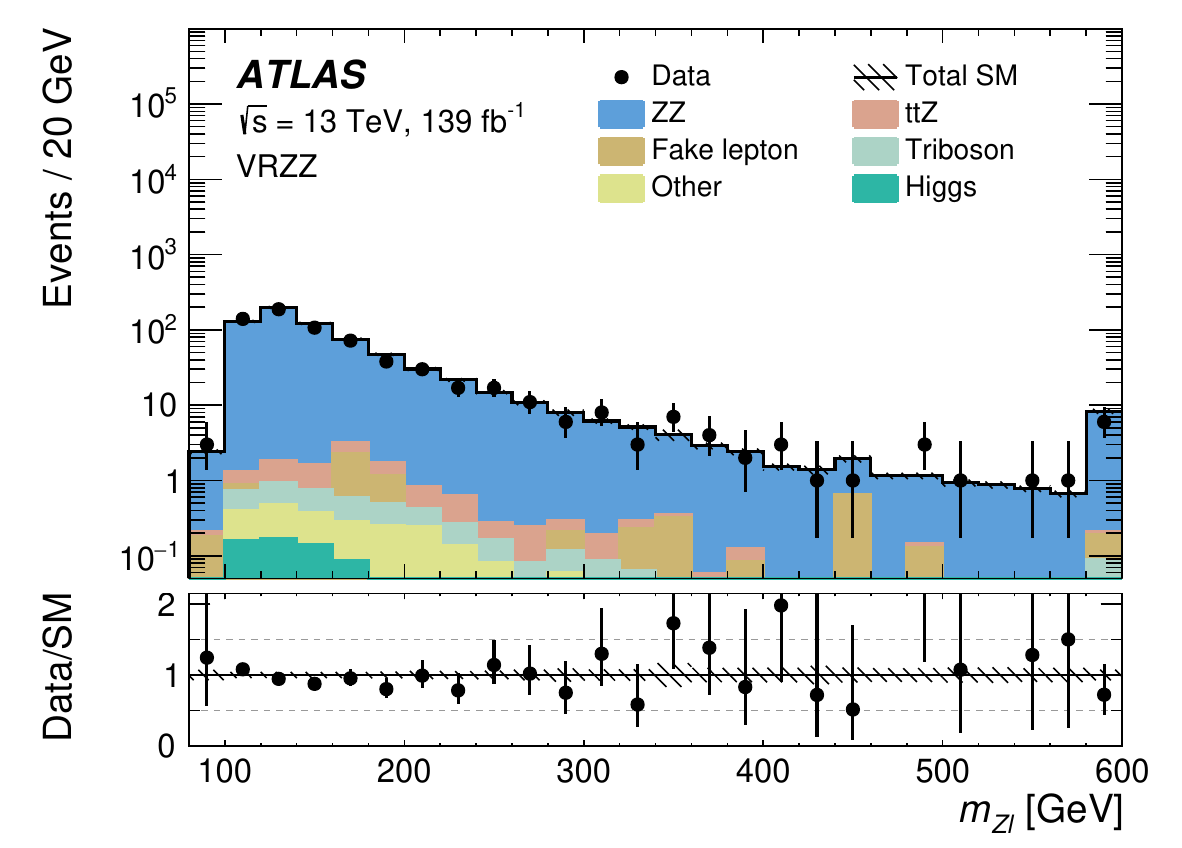}
\includegraphics[width=0.399\textwidth]{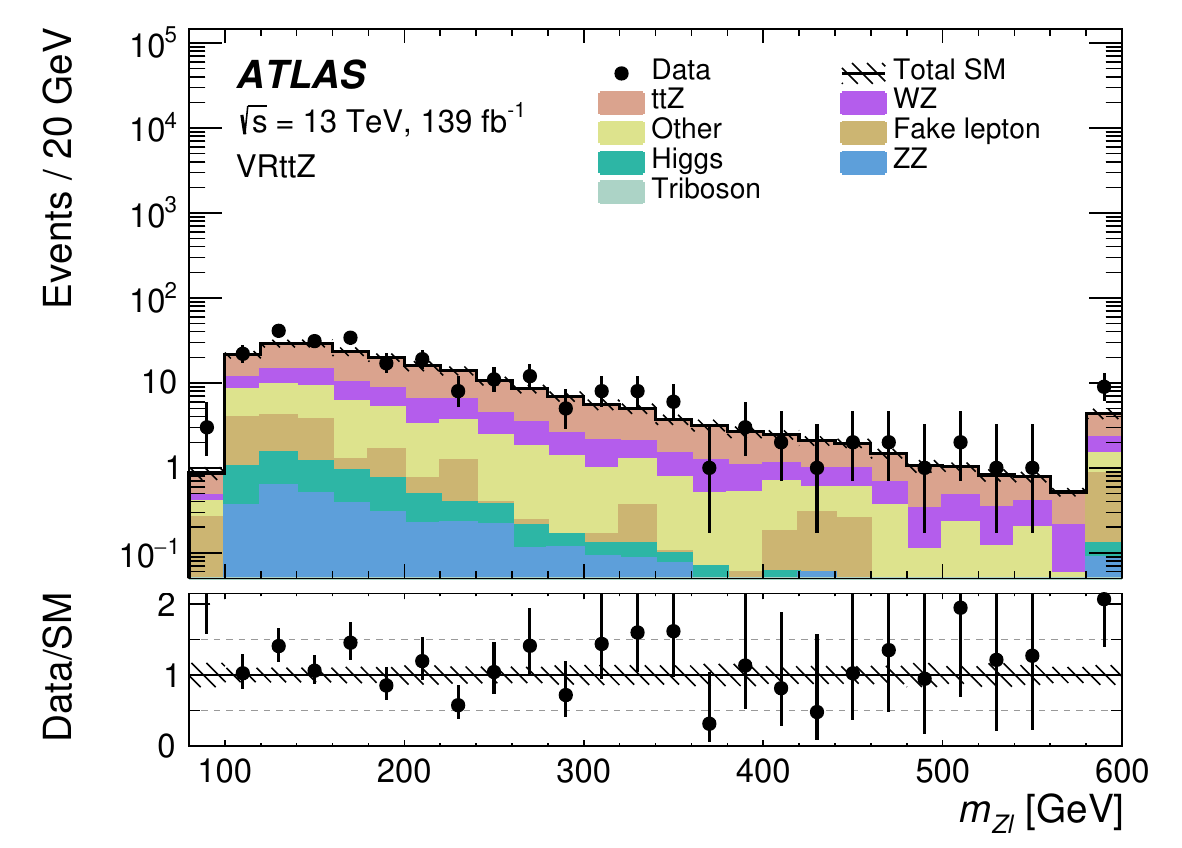}
\includegraphics[width=0.399\textwidth]{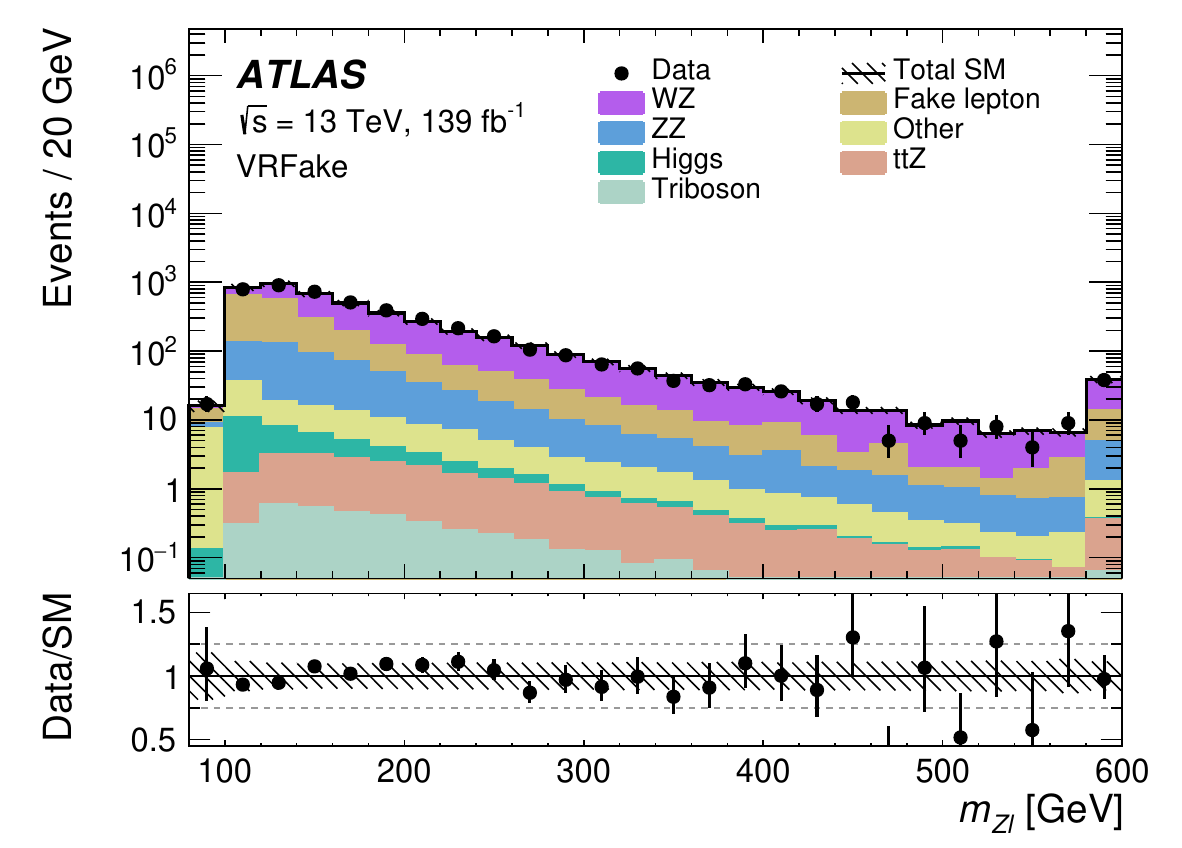}
 
\caption{
\small{
The \mZl distributions of the data and post-fit background in the (from top left to bottom right) \CRWZ, \CRZZ, \CRttZ, \VRmet, \VRmTmin, \VRZZ, \VRttZ, and \VRZj regions.
\captionoverflownounderflow
\captionlegend
\captionsysband
\captionratio
}
}
\label{fig:mZl}
\end{figure}
 
The \ZZ and \ttZ processes are dominant in the four-lepton \SRFour and \SRTL regions.
A control region for the fully leptonic decay in the \ZZ process, \CRZZ, is developed by requiring the presence of a second SFOS pair of electrons or muons with an invariant mass \mllTwo within 5~\GeV of the $Z$ mass.
The \VRZZ validation region has a similar selection, but requires \mllTwo to be between 5 and 20~\GeV of the $Z$ mass, falling naturally between the \CRZZ requirement and the 20~\GeV \mllTwo veto of \SRFour and \SRTL.
The \mllTwo distribution that includes both \CRZZ and \VRZZ is shown in Figure~\ref{fig:CRVR}, and good agreement is seen between data and the post-fit background estimates.
Events for which one $Z$ decays into a pair of $\tau$-leptons that both then subsequently decay leptonically are included in this validation region.
Good modeling in the three-lepton regions is also expected for such \ZZ events when only one $\tau$-lepton decays leptonically,
although this process is strongly suppressed by the \Met and \mTmin requirements.
 
The control region \CRttZ targets the \ttZ process in the SRs, for which the $Z$ boson decays leptonically and one or both top quarks decay leptonically,
and requires at least two $b$-jets in the event.
The \chono may also produce two $b$-jets through the decay of a \Higgs boson, but because of the boost of the Higgs boson they are produced back-to-back less often.
Therefore, the $b$-jets in \CRttZ are required to be produced with $\dRbb > 2.5$, while the SRs require events with at least two $b$-jets to satisfy $\dRbb < 1.5$.
A requirement of $\met > 40~\GeV$ is also imposed to reduce the contamination from the \Zjets process.
To increase the number of events in \CRttZ the lepton multiplicity requirement is relaxed to $N_{\ell} \ge 3$, allowing one top quark to decay fully hadronically.
The presence or absence of a fourth lepton does not bias the other selections as the ratio of three-lepton to four-lepton events in the \ttZ sample is well-modeled.
The \VRttZ validation region is defined with the same selections but requiring $1.5 < \dRbb < 2.5$, falling naturally between \CRttZ and the SRs.
The \dRbb distribution for both  \CRttZ and \VRttZ is shown in Figure~\ref{fig:CRVR}.
To maintain orthogonality between the \ttZ regions and the other CRs used in the fit, a requirement of $\dRbb < 1.5$ is applied to all other analysis regions.
 
The \mZl distributions for the CRs and VRs are given in Figure~\ref{fig:mZl}.
No significant shape disagreement is seen between data and MC simulation, validating the modeling of the backgrounds in \mZl.
The normalization in \CRWZ, \CRZZ, and \CRttZ is therefore performed inclusively in \mZl to improve the statistical precision.
 
The observed event yields in the CRs and VRs are compared with the background estimates and are shown in Figure~\ref{fig:region_yields}.
The CRs are shown with the pre-fit background estimates, and the bottom panel shows the relative disagreement, which is subsequently reduced by the fit.
The VRs are shown with the post-fit background estimates, and the bottom panel shows the significance of the disagreement when accounting for all uncertainties.
Both the CRs and SRs are included in the fit, with the \WZ, \ZZ, and \ttZ normalization factors constrained primarily by the CRs due to their high number of events and purity.
The normalization factors of the background-only fit to the CRs and SRs are $1.01\pm0.03$ for the \WZ process, $1.12\pm0.06$ for the \ZZ process, and $1.05\pm0.18$ for the \ttZ process.
 
\begin{figure}[hb]
\centering
\includegraphics[width=0.65\textwidth]{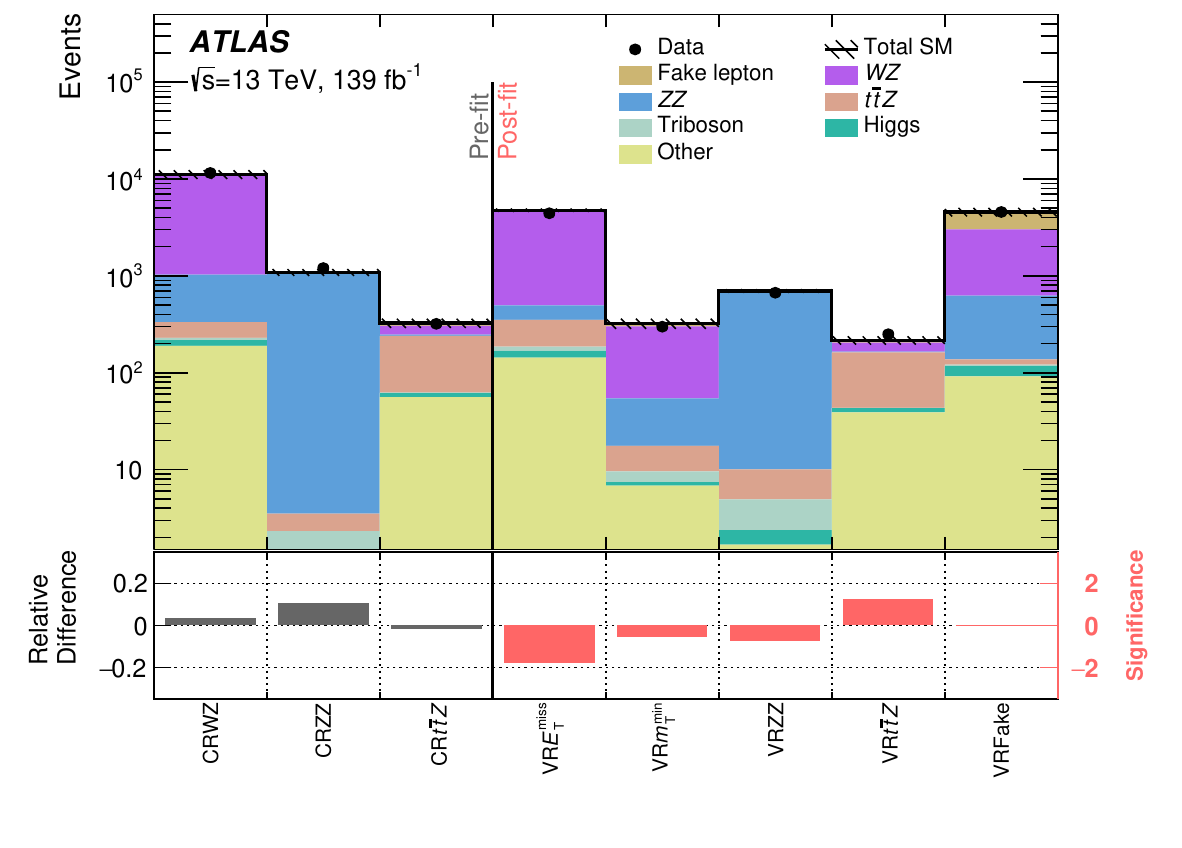}
\caption{The observed data and the SM background expectation in the CRs (pre-fit) and VRs (post-fit).
\captionlegend
\captionsysband
The bottom panel shows the fractional difference between the observed data and expected yields for the CRs
and the significance of the difference for the VRs, computed following the profile likelihood method described in Ref.~\cite{Cousins:2007bmb}. }
\label{fig:region_yields}
\end{figure}
 
The data agree well with the post-fit background estimates in all validation regions, giving confidence in the validity of the post-fit background estimation in the SRs.
A slight overestimation of almost $2\,\sigma$ is seen in \VRmet, and no features are seen in the comparison of data and the post-fit background estimates in the \mZl (Figure~\ref{fig:mZl}) or \met (Figure~\ref{fig:CRVR}) distributions of \VRmet.
A minor excess of data over the background estimation of $1.3\,\sigma$ is seen in \VRttZ, and good agreement is seen in the shape of the relevant \mZl (Figure~\ref{fig:mZl}) and \dRbb (Figure~\ref{fig:CRVR}) distributions.
 
\subsection{Backgrounds from \fake leptons}
\label{sec:background:Fake}
 
Processes that include one or more \fake leptons are estimated with the data-driven \fake-factor method~\cite{STDM-2011-24,ATL-PHYS-PUB-2010-005},
avoiding a reliance on MC simulation to model the prompt-lepton quality criteria of \fake leptons.
The modeling is also made difficult by the many sources of \fake-lepton processes, each of which is kinematically different
and provides a relative contribution to the background estimate that is dependent on the analysis phase-space.
The most relevant sources for this analysis include the in-flight decays of heavy-flavor hadrons (HF) and misidentified light-flavor jets or in-flight decays of pions and kaons (LF).
The \fake muons in this analysis are predominantly from HF sources while \fake electrons are produced from both HF and LF sources,
with their relative contribution varying from 2:1 to 1:5 depending upon the analysis region.
The pair production of two electrons from the conversion of a prompt photon (Conv) is also considered a \fake-lepton process but makes a minor contribution.
In this analysis the relevant \fake processes (and their sources) are \Zjets (LF, HF) and \ttbar (HF) in the three-lepton regions
and \WZ (LF) and \ZZ (LF, Conv) in the four-lepton regions, with \SRTL also having a large contribution from \ttZ (HF).
 
Pair-produced electrons are not considered as \fake leptons if they are produced from the conversion of bremsstrahlung from a prompt electron,
such as that from a leptonically decaying $Z$ boson.
Events with such electrons are not targeted by the \fake-factor method but are instead taken directly from MC simulation, which is considered to adequately model such processes.
These events are included in the Other category and are a minor contribution in \CRWZ and the \fake measurement and validation regions, described below, and are negligible in all other regions.
 
A \fake measurement region \CRZj is designed to target the \Zjets process to provide a selection of events enhanced with \fake leptons from sources representative of those expected in the SRs.
The \CRZj region is not directly included in the fit, but is used to derive the \fake-lepton estimation in each analysis region.
Events are selected by requiring two signal leptons that form an SFOS pair and with an invariant mass within 10~\GeV of the $Z$ boson mass.
One of the two signal leptons is required to have fired a single-lepton trigger, thus ensuring no selection bias from \fake leptons.
To enhance the \Zjets purity and reduce prompt-lepton event contamination from the \WZ process, \CRZj requires $\met<30~\GeV$ and $\mT<30~\GeV$.
A third, unpaired baseline lepton is also required in the event and is designated as the \fake candidate.
A  requirement on the trilepton invariant mass of $\mThreel > 105~\GeV$ reduces contamination from the $Z\rightarrow4\ell$ process.
 
For all regions, events are split into two populations according to whether the \fake candidate meets the nominal signal-quality criteria (nom-ID)
or fails to meet at least one of the signal-lepton identification, isolation, or impact parameter criteria (anti-ID).
The expected contamination by prompt-lepton events from \WZ and \ZZ processes, as estimated from MC simulation, is subtracted from both populations
so that they better represent the yields from \fake-lepton sources.
The \fake factor is defined as the ratio of the yield of nom-ID to anti-ID events in \CRZj and reflects the relative likelihood for a \fake lepton that meets
the baseline criteria to either meet or fail to meet the signal-lepton quality criteria.
This ratio has a dependence on the \fake-lepton source but is fairly independent of the underlying physics process or any additional activity in the event.
Therefore, in each analysis region the \fake factor can be applied to a population of anti-ID events,
defined with the same region selections but with one or more signal leptons replaced by anti-ID leptons,
to predict the yield of \fake-lepton events that have passed the selection requirements.
 
The \fake factors are derived separately for electron and muon \fake candidates and are parameterized as a function of \ptcone,
which better reflects the \pt of the underlying particle that has produced the \fake lepton, such as a HF hadron.
Additional parameterizations of the \fake factor were considered, including lepton $\eta$, \met, and the \bjet multiplicity of the event,
but a two-dimensional parameterization would significantly reduce the statistical precision of the \fake factors.
Alternative parameterizations are instead used to define a systematic uncertainty due to the choice of \ptcone.
The statistical uncertainty of each \fake factor is propagated to an uncertainty in the yield.
An uncertainty due to the prompt-lepton subtraction is estimated by varying the subtracted yields of the \WZ and \ZZ MC simulations up and down by 5\%, corresponding to their cross-section uncertainties~\cite{STDM-2018-03}.
For any \mZl bin of an SR that does not have an anti-ID event, and therefore has a prediction of zero \fake-lepton events,
an uncertainty is applied corresponding to a yield of 0.32 \fake events.
This represents the largest \fake estimate possible given a $1\,\sigma$ upward fluctuation in the anti-ID event yield.
 
To validate the \fake estimation, a dedicated validation region \VRZj is developed closer to the SRs, using the same selections as \CRZj but requiring $\met<40~\GeV$ and $30<\mT<50~\GeV$.
Good agreement is seen between data and the post-fit background estimate in \VRZj, and for the other VRs,
for all observables relevant for the \fake factor, including the \mZl distributions shown in Figure~\ref{fig:mZl}.
A conservative closure uncertainty of 23\% (27\%) is applied to the yield of events with electron (muon) \fake candidates,
and is derived so as to cover the most discrepant \ptcone bin observed in \VRZj.
 
The \fake factor for electrons is sensitive to the relative composition of the \fake sources, which primarily varies between LF and HF in the analysis regions.
To derive an uncertainty in the \fake-source composition, the MC \fake factors are measured in MC simulation in \CRZj for HF and LF sources separately.
The inclusive MC \fake factors are seen to be reproduced by reweighting the HF and LF MC \fake factors according to the \CRZj composition.
Therefore, a composition systematic uncertainty is derived in each analysis region by comparing the inclusive \CRZj MC \fake factors with those
calculated from a reweighting of HF and LF MC \fake factors, according to the composition of that region.
The systematic uncertainty is derived using only MC simulation, in order to provide clean sources of HF and LF \fake electrons,
but is applied to the nominal data-driven \fake factors, and is measured to be at most 53\% for the electron \fake factors in \SRFour.
 
 
\section{Systematic uncertainties}
\label{sec:systematics}
Uncertainties in the expected signal and background yields account for the statistical uncertainties of the MC samples,
the experimental systematic uncertainties in the detector measurements, and the theoretical systematic uncertainties of the MC simulation modeling.
The uncertainties of the major backgrounds normalized in the CRs reflect the limited statistical precision of the CRs
and the systematic uncertainties in the extrapolation to the signal regions,
and an additional uncertainty in the normalization factor from the combined fit is included.
The uncertainties related to the data-driven \fake background estimation are described in detail in Section~\ref{sec:background:Fake}.
 
Systematic uncertainties are treated as Gaussian nuisance parameters in the likelihood while the statistical uncertainties of the MC samples are treated as Poisson nuisance parameters.
Unless stated otherwise, each experimental uncertainty is treated as fully correlated across the analysis regions, while each theoretical uncertainty is derived as the
relative yield between an analysis region and a control region and is treated as uncorrelated across analysis regions.
 
\begin{figure}[h]
\centering
\includegraphics[width=0.48\textwidth]{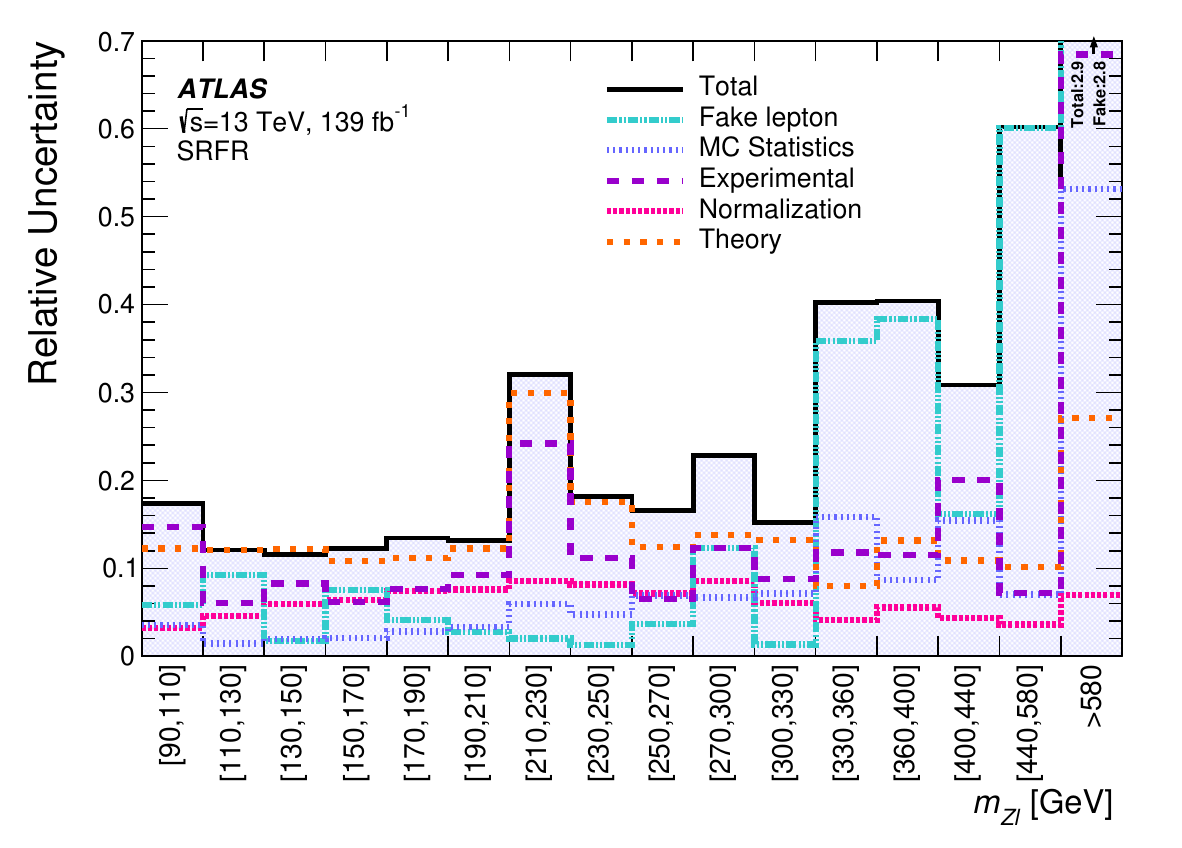}
\includegraphics[width=0.48\textwidth]{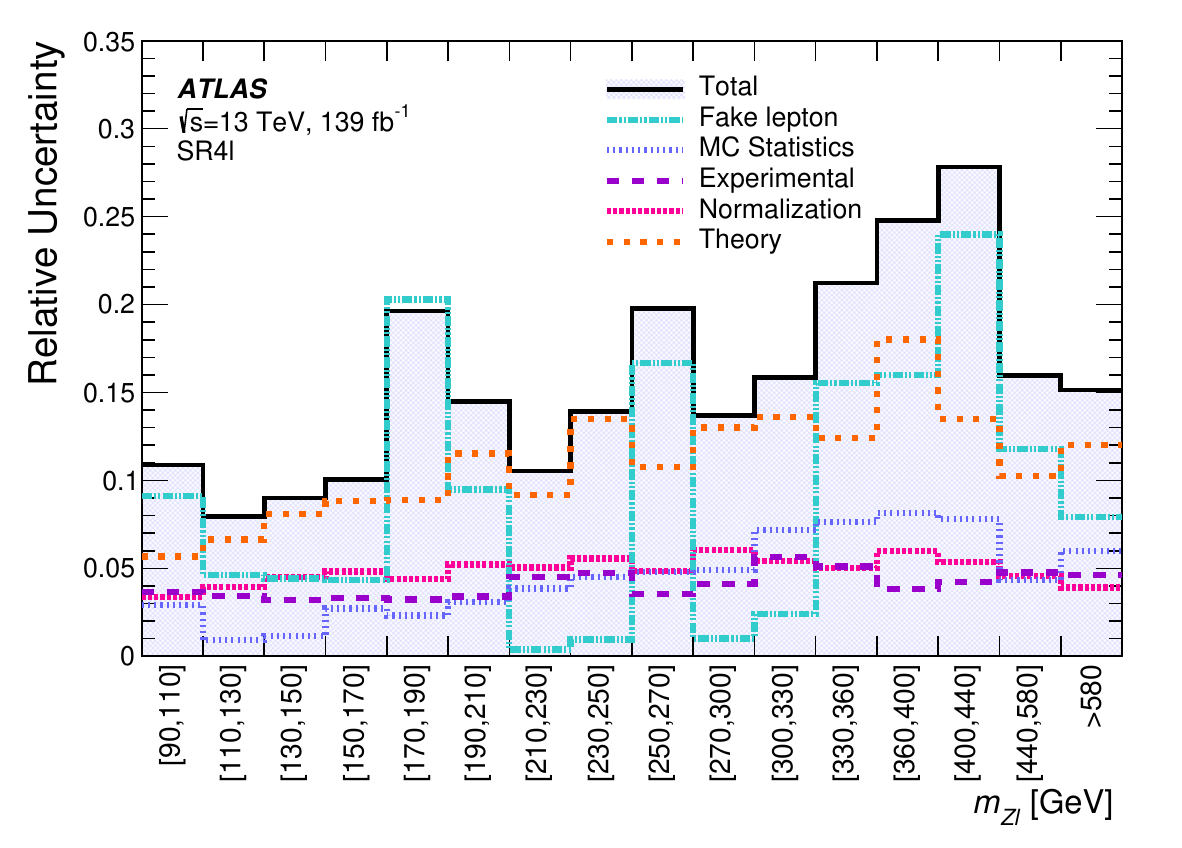}
\includegraphics[width=0.48\textwidth]{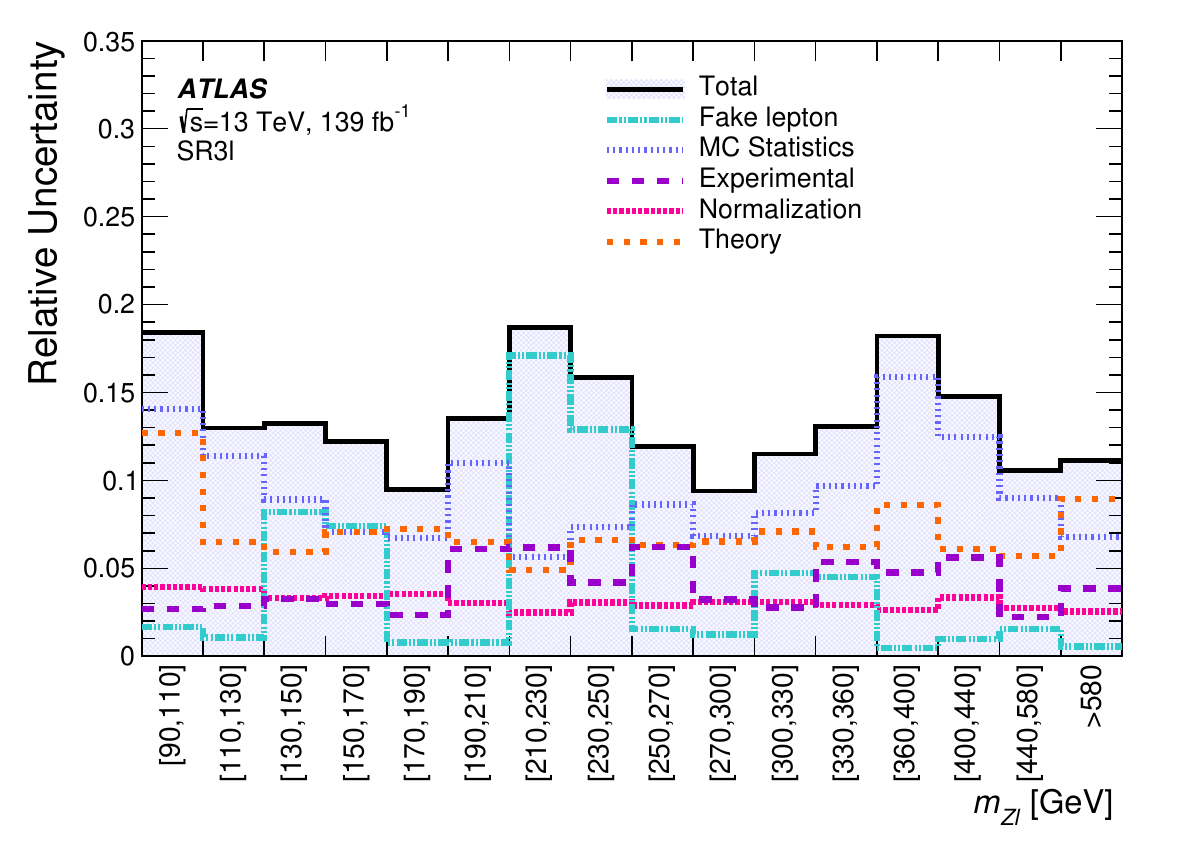}
\caption{
The relative uncertainties in the post-fit SM background prediction as a function of \mZl from the background-only
fit for the (top left) \SRTL, (top right) \SRFour, and (bottom) \SRThree regions.
The \mZl binning (Eq.~(\ref{eq:binning})) is the same as that used in the fit.
Sources of uncertainty are grouped into experimental, theoretical, and MC statistical categories.
Separate categories are provided for the \fake backgrounds and for the normalization procedure of the major \WZ, \ZZ, and \ttZ backgrounds.
The individual uncertainties can be correlated and do not necessarily contribute in quadrature to the total uncertainty.}
\label{fig:systematics}
\end{figure}
 
A summary of the background uncertainties is shown in Figure~\ref{fig:systematics}.
Individual uncertainties can be correlated or anti-correlated, for example between an uncertainty on a major background and the uncertainty on the CR-to-SR normalization procedure for that background.
Bin-to-bin fluctuations in the uncertainty of the \fake background estimation reflect the small anti-ID population and the conservative uncertainties applied when no anti-ID events are seen in the data.
The effect of localized fluctuations in one SR is limited as all three SRs contribute to the overall sensitivity.
A relative uncertainty of 2.9 is seen in the last \mZl bin of \SRTL and is driven by a relative uncertainty of 2.8 in the \fake estimation, reflecting the small post-fit background expectation.
 
Experimental uncertainties in the detector measurements reflect the accuracy of the kinematic measurements of jets, electrons,
muons, and \met.
Varying the scale or resolution of the energy or \pt of objects within the uncertainties can cause the migration of events between \mZl bins or affect the inclusion of an event in an analysis region.
The jet energy scale and resolution uncertainties~\cite{PERF-2016-04,PERF-2011-04} are a large component of the experimental uncertainty.
They are derived as a function of jet \pt and $\eta$ and account for the flavor and pileup dependencies of the detector energy measurement.
Similar scale and resolution uncertainties are included for electrons~\cite{EGAM-2018-01} and muons~\cite{PERF-2015-10}.
These per-object uncertainties are propagated through the \met calculation, with additional uncertainties accounting for the scale and resolution of the \met soft term~\cite{PERF-2016-07}.
 
Additional experimental uncertainties account for the mismodeling in MC simulation of observables related to the detection of leptons and jets.
They include the efficiency of the triggering, identification, reconstruction, and isolation requirements of electrons~\cite{EGAM-2018-01} and muons~\cite{PERF-2015-10}.
They also include the identification and rejection of pileup jets by the jet vertex tagger~\cite{PERF-2014-03}
and the identification of $b$-jets by the flavor-tagging algorithm~\cite{FTAG-2018-01}.
The experimental uncertainty in the combined 2015--2018 integrated luminosity is 1.7\%~\cite{ATLAS-CONF-2019-021},
obtained primarily using the luminosity measurements of the LUCID-2 detector~\cite{LUCID2}.
 
Theoretical uncertainties in the shape of the major diboson, triboson, and \ttZ backgrounds are derived using MC simulation with varied generator parameters.
For the other minor backgrounds a conservative 20\% uncertainty is assumed.
This value is larger than is typically expected for the minor background processes and the choice has a negligible effect on the final results due to the small contributions of these backgrounds.
Uncertainties due to the choice of QCD renormalization and factorization scales~\cite{Bothmann:2016nao} are assessed by
varying the relevant generator parameters up and down by a factor of two around the nominal values,
allowing for both independent and correlated variations of the two scales but prohibiting anti-correlated variations.
Each QCD variation is kept separate and is treated as correlated across analysis regions.
An uncertainty of 1\% due to the chosen value of the strong coupling constant \alphas is assessed by varying \alphas by $\pm 0.001$ in the generator parameter settings.
Uncertainties related to the choice of PDF sets, CT14NNLO~\cite{Dulat:2015mca} or MMHT2014NNLO~\cite{Harland-Lang:2014zoa},
are derived by taking the envelope of the variation in event yield of 100 propagated uncertainties~\cite{Butterworth:2015oua}.
 
Additional theoretical uncertainties are assessed for the major backgrounds.
These are related to assumptions made in the event generators and PS models,
which can affect both the event kinematics and the cross section of the physics process.
For the diboson backgrounds, the \sherpa parameters related to the PS matching scale and resummation scale are varied up and down by a factor of two around the nominal values,
and an alternative recoil scheme is studied.
For the \ttZ background, the uncertainties in the hard scatter and in the PS are derived through a comparison with
the \sherpa and \amchseven predictions, respectively.
Additional uncertainties in the amount of initial-state radiation (ISR) in the \ttZ background are assessed by varying the
related generator parameters.
 
For the signal samples, theoretical uncertainties in the cross section are applied, ranging from 4.5\% at 100~\GeV to 16\% at 1500~\GeV.
Uncertainties related to the QCD scale, PS matching scale, and amount of ISR
are derived by varying the related generator parameters of the A14 tune~\cite{ATL-PHYS-PUB-2014-021}.
 
 
\section{Results}
\label{sec:result}
 
The data are compared with the post-fit background expectations, derived from a background-only profile likelihood fit of all CRs and SRs simultaneously as described in Section~\ref{sec:background},
and no significant excess is observed.
The VRs, shown previously in Figure~\ref{fig:region_yields}, demonstrate good modeling of the post-fit background expectation in regions kinematically similar to the SRs
and for a variety of observables, validating the background-estimation technique.
The observed and expected numbers of events in \SRTL, \SRFour, and \SRThree  are given in
Table~\ref{tab:inclusive_SR_yields}
inclusively in \mZl and for the inclusive, $e$ direct-lepton, and $\mu$ direct-lepton flavor channels.
The background expectation and uncertainty are further split into contributions from each category of SM processes.
Separate fits are performed for each flavor channel and for the inclusive channel, and therefore the predicted yields in the $e$ and $\mu$ channels may not necessarily add to the inclusive yield.
Additionally, the \SRTL regions have the same flavor requirement on both direct leptons in an event, and the data and predicted yields in the $e$ and $\mu$ channels do not add to the inclusive result.
 
\begin{table}
\caption{
The observed yields and post-fit background expectations in \SRTL, \SRFour, and \SRThree,
shown inclusively and when the direct lepton from a \chono decay is required to be an electron or muon.
\captionlegend
Uncertainties in the background expectation include combined statistical and systematic uncertainties.
The individual uncertainties may be correlated and do not necessarily combine in quadrature to give the total background uncertainty.
}
\label{tab:inclusive_SR_yields}
\begin{center}
\setlength{\tabcolsep}{0.0pc}
{\small
\begin{tabular*}{\textwidth}{@{\extracolsep{\fill}}lcccccc}
\noalign{\smallskip}\hline\noalign{\smallskip}
{Region}                    & \SRTL                            & \SRTLE                           & \SRTLM                           & \SRFour                             & \SRFourE                            & \SRFourM   \\
\noalign{\smallskip}\hline\noalign{\smallskip}
Observed yield             & $42$                                & $15$                                & $17$                                & $89$                                & $48$                                & $41$   \\
\noalign{\smallskip}\hline\noalign{\smallskip}
Expected background yield  & $\phantom{}39 \pm 4\phantom{0}$ & $\phantom{}13.7 \pm 2.0\phantom{0}$ & $\phantom{}15.7 \pm 2.5\phantom{0}$ & $\phantom{.0}76 \pm 6\phantom{.00}$ & $\phantom{.0}35.8 \pm 3.5\phantom{.00}$ & $\phantom{}38.2 \pm 2.8\phantom{0}$ \\
\noalign{\smallskip}\hline\noalign{\smallskip}
$WZ$ yield                 & $-$                                 & $-$                                 & $-$                                 & $-$                                 & $-$                                 & $-$         \\
$ZZ$ yield                 & $\phantom{}19 \pm 4\phantom{0}$     & $7.1 \pm 1.7$                       & $\phantom{}10.4 \pm 2.4\phantom{0}$ & $\phantom{}20.9 \pm 1.1\phantom{0}$ & $9.5 \pm 0.6$                       & $\phantom{}11.2 \pm 0.7\phantom{0}$ \\
$t\bar{t}Z$ yield          & $\phantom{}12.2 \pm 3.2\phantom{0}$ & $2.4 \pm 0.7$                       & $3.0 \pm 0.6$                       & $\phantom{.0}18 \pm 6\phantom{.00}$ & $9.1 \pm 3.2$                       & $8.5 \pm 1.6$         \\
Triboson yield             & $1.3 \pm 0.4$                       & $0.25 \pm 0.09$                     & $0.33 \pm 0.12$                     & $\phantom{}12.2 \pm 2.8\phantom{0}$ & $5.8 \pm 1.4$                       & $6.0 \pm 1.5$  \\
Higgs yield                & $2.6 \pm 0.5$                       & $0.72 \pm 0.17$                     & $1.17 \pm 0.25$                     & $\phantom{}11.2 \pm 2.0\phantom{0}$ & $5.3 \pm 1.0$                       & $5.5 \pm 1.1$ \\
Other yield                & $2.1 \pm 0.5$                       & $0.25 \pm 0.17$                     & $0.39 \pm 0.16$                     & $7.9 \pm 1.5$                       & $4.0 \pm 0.8$                       & $3.5 \pm 0.8$          \\
\emph{Fake} yield          & $1.3 \pm 0.8$                       & $3.0 \pm 1.5$                       & $0.5_{-0.5}^{+0.6}$~                 & $6.4 \pm 2.5$                       & $2.1 \pm 1.1$                       & $3.6 \pm 1.7$          \\
\noalign{\smallskip}\hline\noalign{\smallskip}
{Region}                   & \SRThree                            & \SRThreeE                           & \SRThreeM                              & & & \\[-0.05cm]
\noalign{\smallskip}\hline\noalign{\smallskip}
Observed yield             & $61$                                & $28$                                & $33$                                & & & \\
\noalign{\smallskip}\hline\noalign{\smallskip}
Expected background yield  & $\phantom{}54.9 \pm 3.3\phantom{0}$ & $\phantom{}27.5 \pm 2.2\phantom{0}$ & $\phantom{}27.4 \pm 2.0\phantom{0}$ & & & \\
\noalign{\smallskip}\hline\noalign{\smallskip}
$WZ$ yield                 & $\phantom{}33.6 \pm 2.4\phantom{0}$ & $\phantom{}16.5 \pm 1.7\phantom{0}$ & $\phantom{}17.3 \pm 1.8\phantom{0}$ & & & \\
$ZZ$ yield                 & $0.92 \pm 0.27$                     & $0.11 \pm 0.04$                     & $0.77 \pm 0.24$                     & & & \\
$t\bar{t}Z$ yield          & $7.5 \pm 2.3$                       & $4.1 \pm 1.3$                       & $3.4 \pm 0.7$                       & & & \\
Triboson yield             & $5.6 \pm 1.5$                       & $2.7 \pm 0.8$                       & $2.6 \pm 0.7$                       & & & \\
Higgs yield                & $0.51 \pm 0.10$                     & $0.25 \pm 0.06$                     & $0.23 \pm 0.05$                     & & & \\
Other yield                & $4.2 \pm 0.8$                       & $2.0 \pm 0.4$                       & $2.0 \pm 0.4$                       & & & \\
\emph{Fake} yield          & $2.5 \pm 1.2$                       & $1.8 \pm 1.1$                       & $1.0 \pm 0.8$                       & & & \\
\noalign{\smallskip}\hline\noalign{\smallskip}
\end{tabular*}
}
\end{center}
 
\end{table}
 
The \mZl distributions in each SR, with binning corresponding to that used in the fit, are shown in Figure~\ref{fig:SR_yield}.
The SRs show good agreement in the shape of the \mZl distribution between data and the SM expectation, with no significant localized excesses.
Three example signals of mass 200, 500, and 800~\GeV are included in these figures and peak strongly in their target \mZl bin for all three SRs,
with the 800~\GeV signal only visible in the last \mZl bin.
Other observables in the SRs relevant for the extrapolation of the yield normalization are shown in Figure~\ref{fig:SR_dist} and also demonstrate good agreement.
 
\begin{figure}[hb!]
\centering
\includegraphics[width=0.48\textwidth]{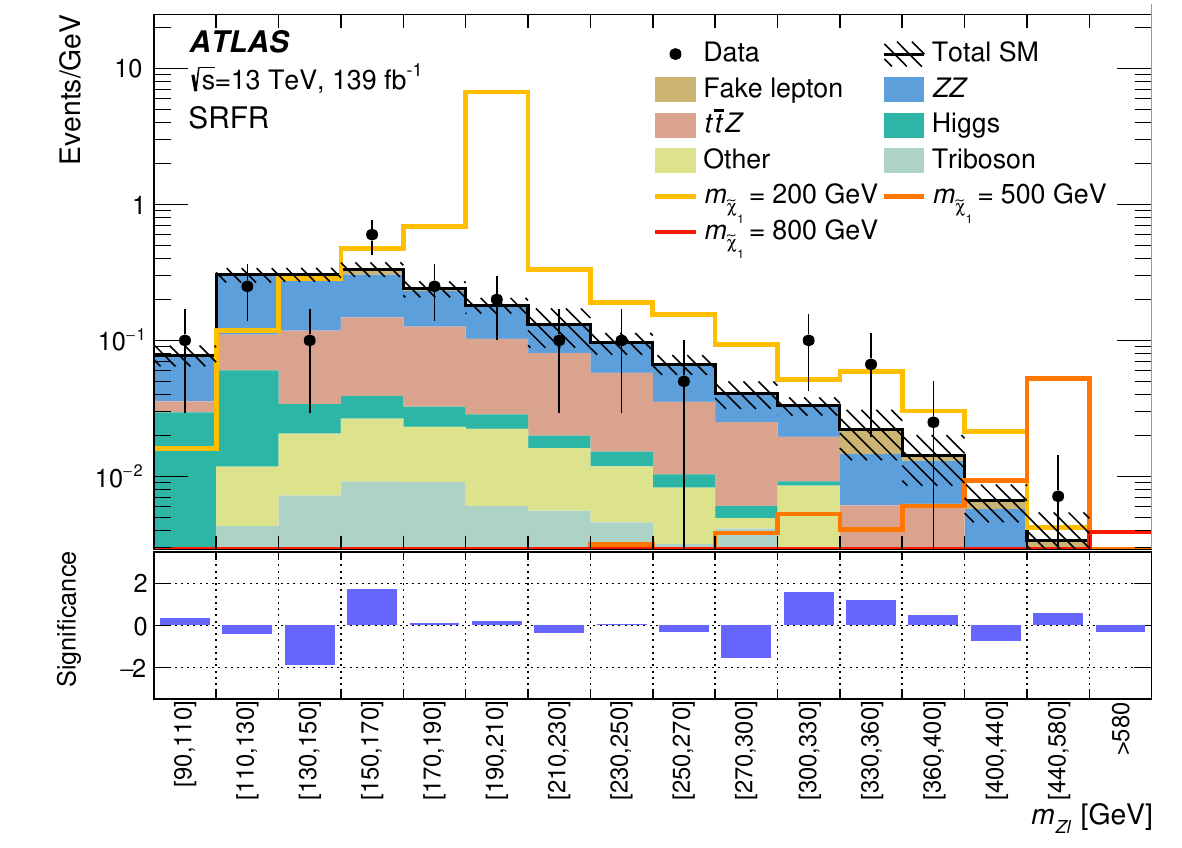}
\includegraphics[width=0.48\textwidth]{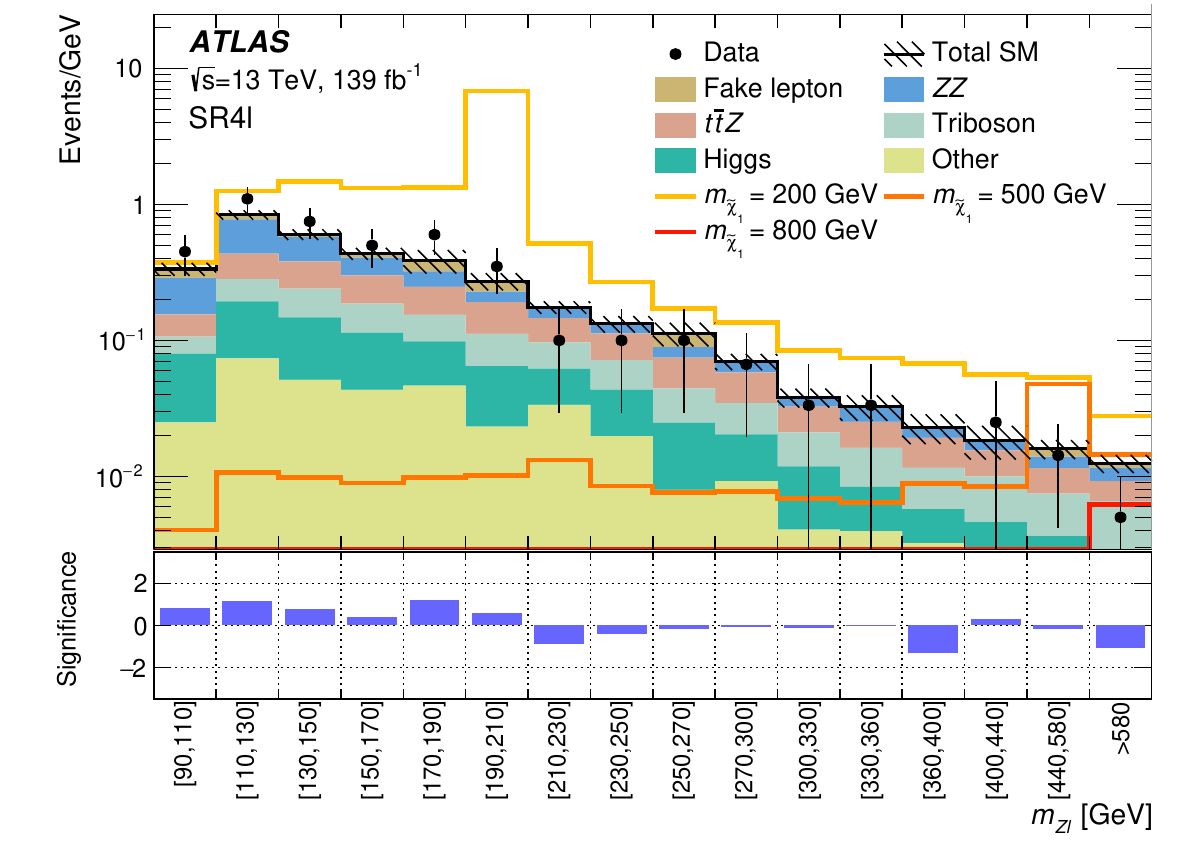}
\includegraphics[width=0.48\textwidth]{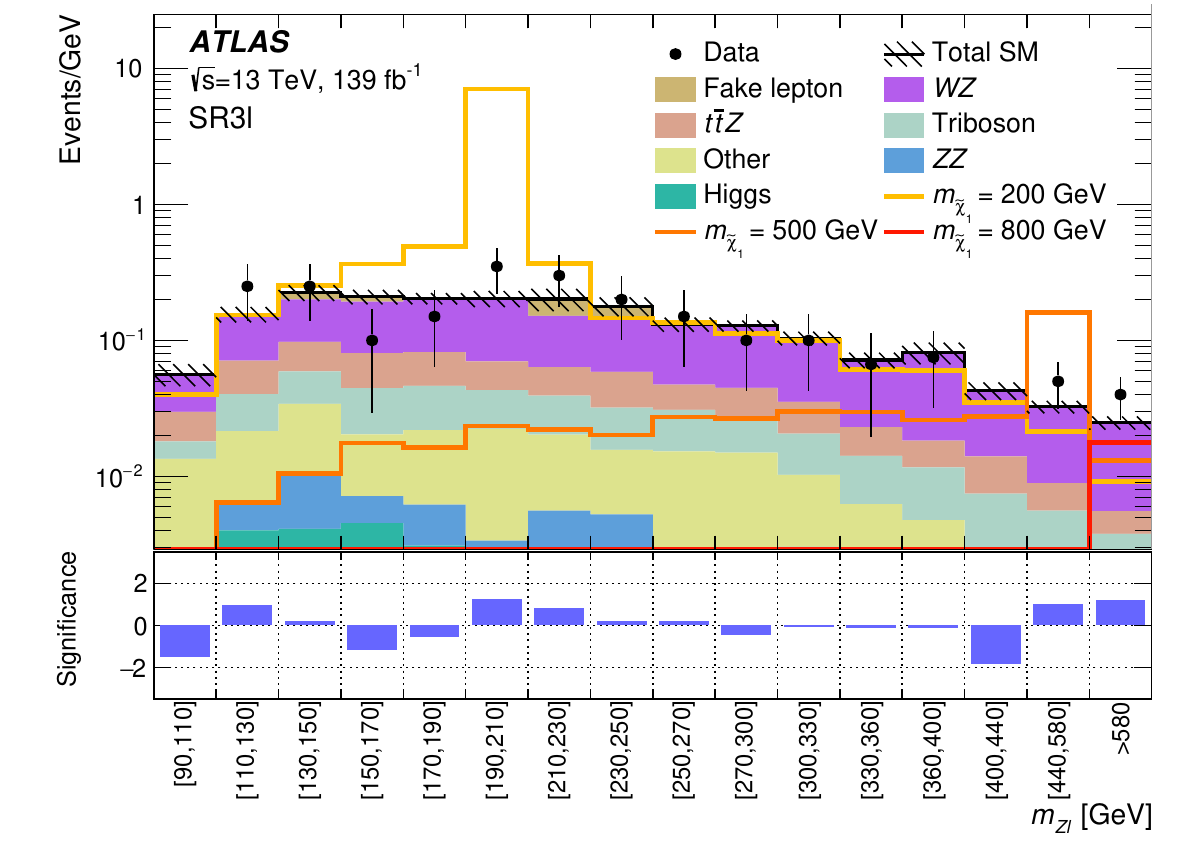}
\caption{The observed data and post-fit SM background expectation as a function of \mZl in
(top left) \SRTL, (top right) \SRFour, and (bottom) \SRThree.
The \mZl binning (Eq.~(\ref{eq:binning})) is the same as that used in the fit and the yield is normalized to the bin width, with the last bin normalized using a width of 200~\GeV.
\captionlegend
\captionsysband
The bottom panel shows the significance of the differences between the observed data and expected yields,
computed following the profile likelihood method described in Ref.~\cite{Cousins:2007bmb}. }
 
\label{fig:SR_yield}
\end{figure}
 
\begin{figure}[hb!]
\centering
\includegraphics[width=0.48\textwidth]{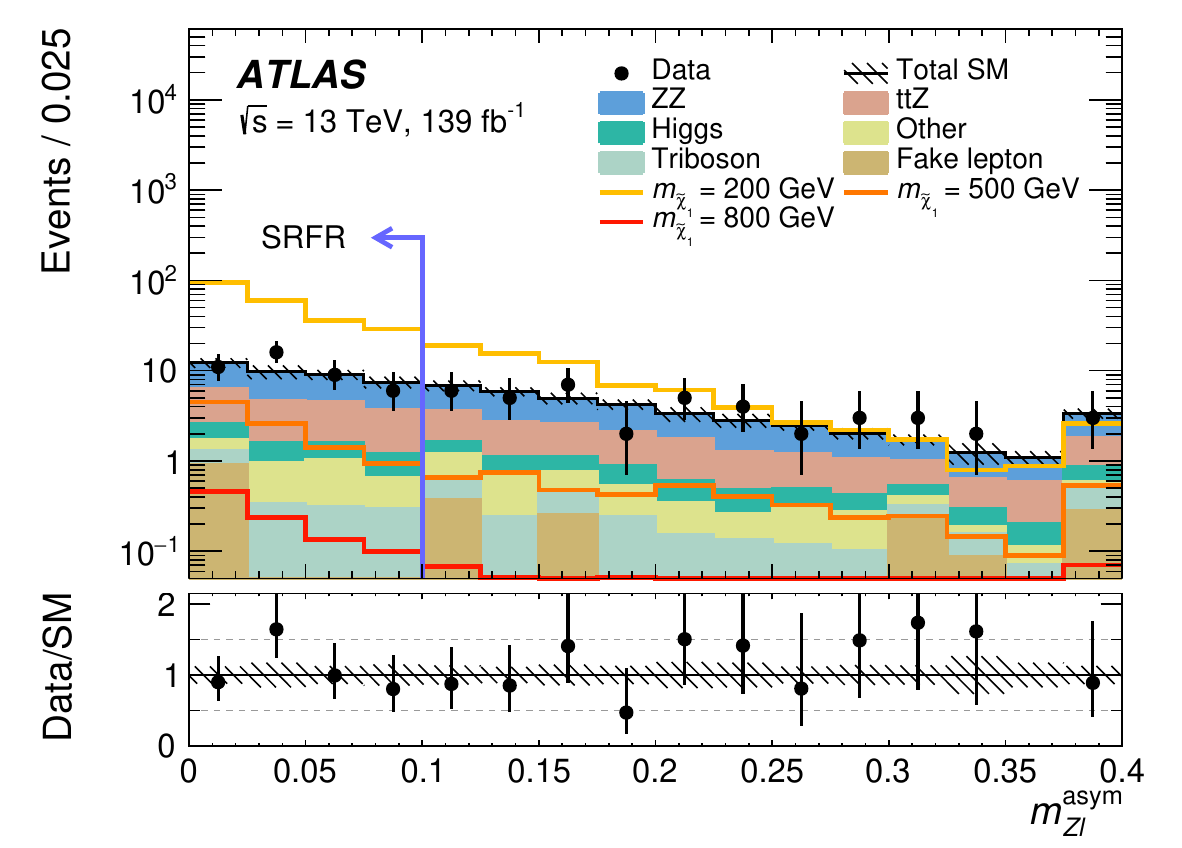}
\includegraphics[width=0.48\textwidth]{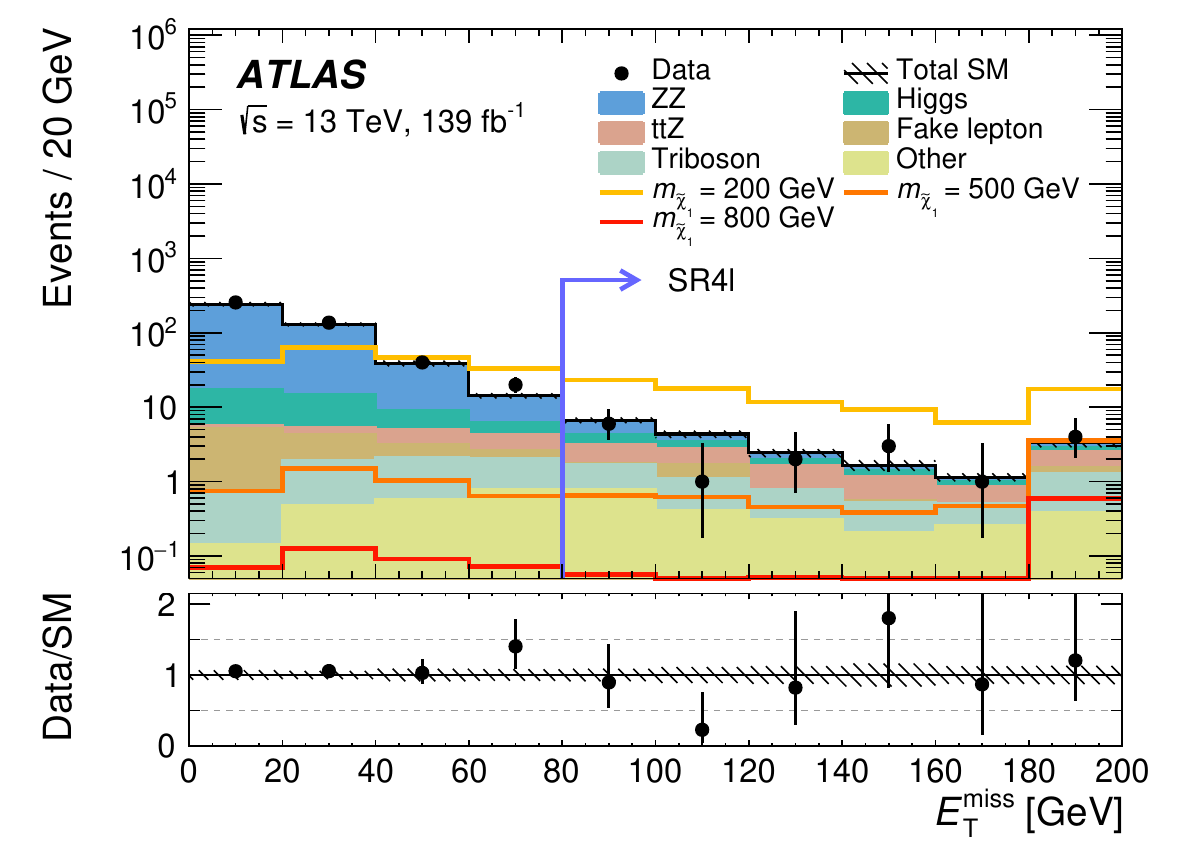}
\includegraphics[width=0.48\textwidth]{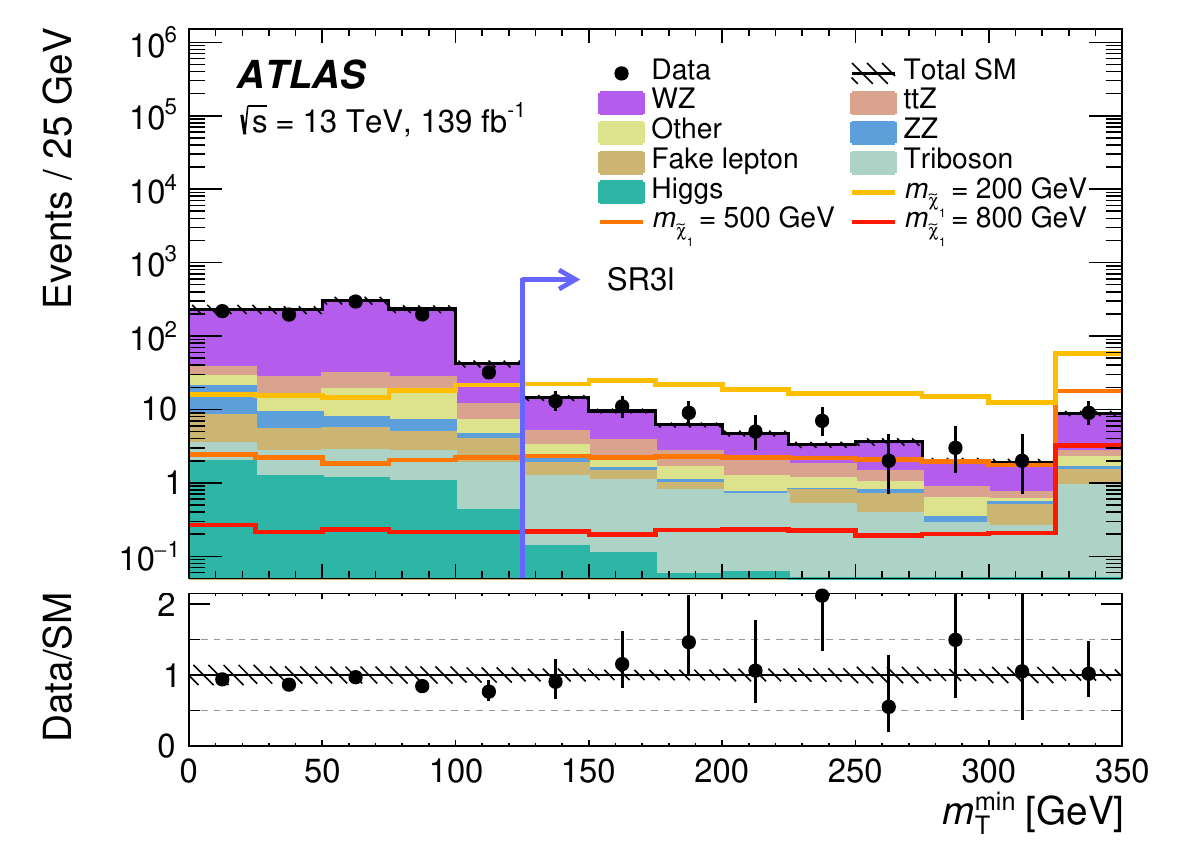}
\includegraphics[width=0.48\textwidth]{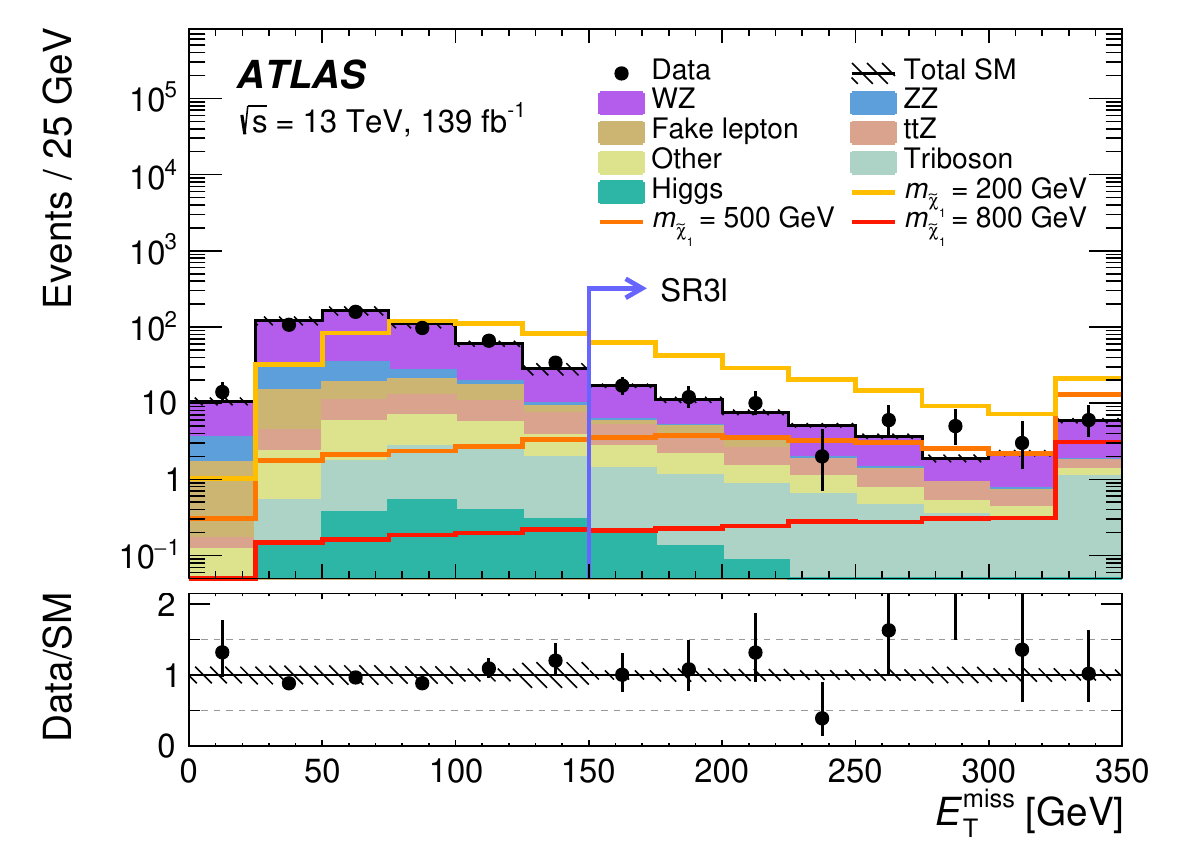}
\caption{ Example kinematic distributions in the signal regions showing the data and the post-fit background expectation, including
(top left) \mZlAsym in \SRTL, (top right) \met in \SRFour, and
(bottom left) \mTmin and (bottom right) \met in \SRThree.
The fit uses all CRs and SRs, and the distributions are shown inclusively in \mZl.
The full event selection for each of the corresponding regions is applied except for the variable shown, where the selection is indicated by a blue arrow.
\captionoverflownounderflow
\captionlegend
\captionsysband
\captionratio
}
 
\label{fig:SR_dist}
\end{figure}
 
\FloatBarrier

\subsection{Model-independent limits on new physics in inclusive regions}
 
Upper limits are set on the possible visible cross sections of generic beyond-the-SM (BSM) processes in each \mZl bin of each SR.
These model-independent limits are derived at 95\% confidence level (CL) using the \CLs prescription~\cite{Read:2002hq},
and results are evaluated using pseudo-experiments.
A profile likelihood fit is performed on the numbers of observed and expected events in the target \mZl bin of one SR and the three CRs,
and a generic BSM process is assumed to contribute only to the target \mZl bin.
In this way no assumption is made concerning the \chono branching fractions or \mZl shape of the BSM process.
No uncertainties in the yield of the BSM process are considered, except for the luminosity uncertainty.
 
This procedure is repeated for each of the 16 \mZl bins in each of the three SRs, with only one SR bin considered for each fit.
This differs from the nominal fit strategy which is performed using the three CRs and the 48 \mZl bins of the SRs simultaneously,
and minor differences from the significances shown in the bottom panel of Figure~\ref{fig:SR_yield} are seen.
 
\begin{table}[ht!]
\caption{
\small{
Model-independent results where each row targets one \mZl bin of one SR and probes scenarios where a generic beyond-the-SM process is assumed to contribute only to that \mZl bin.
The first two columns refer to the signal region and \mZl bin probed, while the third and fourth columns show the observed (\nObs) and expected (\nExp) event yields.
The expected yields are obtained using a background-only fit of all the CRs, and the errors include statistical and systematic uncertainties.
The fifth and sixth columns show the observed 95\% CL upper limit on the visible cross section (\visXS) and on the number of signal events (\SObs),
while the seventh column shows the expected 95\% CL upper limit on the number of signal events (\SExp) with the associated $1\,\sigma$ uncertainties.
The last column provides the discovery $p$-value and significance ($Z$) of any excess of data above background expectation.
Cases for which the observed yield is less than the expected yield are capped at a $p$-value of 0.5.
}
}
\label{tab:discovery}
 
\centering
\setlength{\tabcolsep}{0.0pc}
{\scriptsize
\begin{tabular*}{\textwidth}{@{\extracolsep{\fill}}rccccccc}
\noalign{\smallskip}\hline\noalign{\smallskip}
Region & Range of \mZl [GeV] & \nObs & \nExp & \visXS [fb]  &  \SObs  & \SExp & $p(s=0)$ ($Z$)  \\
\noalign{\smallskip}\hline\noalign{\smallskip}
\multirow{16}{*}{\rotatebox[origin=r]{90}{\SRTL}} & ~~[90, 110] & $2$ & $1.58 \pm 0.32$ & $0.03$ & $4.2$ & $ { 4.0 }^{ +1.8 }_{ -0.8 }$ &  0.44 (0.2) \\
& [110, 130] & $5$ & $5.9 \pm 1.0$ & $0.04$ & $5.6$ & $ { 6.7 }^{ +2.3 }_{ -2.1 }$ &  0.50 (0.0) \\
& [130, 150] & $2$ & $6.0 \pm 1.0$ & $0.03$ & $3.8$ & $ { 6.3 }^{ +2.0 }_{ -2.0 }$ &  0.50 (0.0) \\
& [150, 170] & $12$~~ & $6.1 \pm 1.0$ & $0.10$ & $14.3$~~ & $ { 7.8 }^{ +3.1 }_{ -1.9 }$ &  0.02 (2.1) \\
& [170, 190] & $5$ & $4.5 \pm 0.8$ & $0.05$ & $6.8$ & $ { 5.8 }^{ +2.2 }_{ -1.3 }$ &  0.31 (0.5) \\
& [190, 210] & $4$ & $3.4 \pm 0.6$ & $0.04$ & $6.2$ & $ { 5.3 }^{ +2.1 }_{ -1.2 }$ &  0.28 (0.6) \\
& [210, 230] & $2$ & $2.6 \pm 1.5$ & $0.03$ & $4.5$ & $ { 4.8 }^{ +1.9 }_{ -0.9 }$ &  0.50 (0.0) \\
& [230, 250] & $2$ & $1.83 \pm 0.31$ & $0.03$ & $4.7$ & $ { 4.1 }^{ +1.6 }_{ -0.8 }$ &  0.41 (0.2) \\
& [250, 270] & $1$ & $1.25 \pm 0.22$ & $0.03$ & $4.0$ & $ { 3.9 }^{ +1.2 }_{ -0.9 }$ &  0.50 (0.0) \\
& [270, 300] & $0$ & $1.18 \pm 0.32$ & $0.03$ & $3.6$ & $ { 3.9 }^{ +1.3 }_{ -0.8 }$ &  0.50 (0.0) \\
& [300, 330] & $3$ & $0.89 \pm 0.16$ & $0.05$ & $6.7$ & $ { 4.2 }^{ +0.8 }_{ -0.4 }$ &  0.02 (2.0) \\
& [330, 360] & $2$ & $0.52 \pm 0.18$ & $0.04$ & $5.6$ & $ { 3.5 }^{ +0.8 }_{ -0.2 }$ &  0.03 (1.9) \\
& [360, 400] & $1$ & $0.50 \pm 0.19$ & $0.03$ & $4.0$ & $ { 3.2 }^{ +1.0 }_{ -0.1 }$ &  0.18 (0.9) \\
& [400, 440] & $0$ & $0.27 \pm 0.09$ & $0.02$ & $3.2$ & $ { 3.1 }^{ +0.8 }_{ -0.1 }$ &  0.50 (0.0) \\
& [440, 580] & $1$ & $0.29 \pm 0.17$ & $0.03$ & $4.4$ & $ { 3.3 }^{ +1.0 }_{ -0.1 }$ &  0.12 (1.2) \\
& >580 & $0$ & ${0.07}^{ +0.17 }_{ -0.07 }$~ & $0.02$ & $3.0$ & $ { 3.1 }^{ +0.1 }_{ -0.0 }$ &  0.50 (0.0) \\
 
\noalign{\smallskip}\hline\noalign{\smallskip}
\multirow{16}{*}{\rotatebox[origin=r]{90}{\SRFour}}  & ~~[90, 110] & $9$ & $6.1 \pm 0.7$ & $0.07$ & $9.6$ & $ { 7.1 }^{ +2.1 }_{ -1.1 }$ &  0.14 (1.1) \\
& [110, 130] & $22$~~ & $15.4 \pm 1.3$~~ & $0.12$ & $16.7$~~ & $ { 10.0 }^{ +4.4 }_{ -1.9 }$~~ &  0.05 (1.6) \\
& [130, 150] & $15$~~ & $10.9 \pm 0.9$~~ & $0.09$ & $12.8$~~ & $ { 8.4 }^{ +3.4 }_{ -1.8 }$ &  0.09 (1.4) \\
& [150, 170] & $10$~~ & $7.9 \pm 0.8$ & $0.07$ & $10.0$~~ & $ { 7.5 }^{ +2.9 }_{ -1.5 }$ &  0.16 (1.0) \\
& [170, 190] & $12$~~ & $5.9 \pm 0.6$ & $0.10$ & $14.2$~~ & $ { 8.5 }^{ +3.4 }_{ -0.8 }$ &  0.02 (2.0) \\
& [190, 210] & $7$ & $4.9 \pm 0.7$ & $0.06$ & $8.3$ & $ { 6.5 }^{ +1.9 }_{ -1.5 }$ &  0.15 (1.0) \\
& [210, 230] & $2$ & $3.17 \pm 0.33$ & $0.03$ & $4.4$ & $ { 4.9 }^{ +2.2 }_{ -1.4 }$ &  0.50 (0.0) \\
& [230, 250] & $2$ & $2.36 \pm 0.27$ & $0.03$ & $4.5$ & $ { 4.5 }^{ +1.9 }_{ -1.1 }$ &  0.50 (0.0) \\
& [250, 270] & $2$ & $2.1 \pm 0.5$ & $0.03$ & $4.9$ & $ { 4.8 }^{ +1.8 }_{ -1.2 }$ &  0.50 (0.0) \\
& [270, 300] & $2$ & $1.88 \pm 0.21$ & $0.03$ & $4.8$ & $ { 4.2 }^{ +1.6 }_{ -1.0 }$ &  0.50 (0.0) \\
& [300, 330] & $1$ & $1.03 \pm 0.14$ & $0.03$ & $4.1$ & $ { 3.6 }^{ +1.6 }_{ -0.5 }$ &  0.50 (0.0) \\
& [330, 360] & $1$ & $0.88 \pm 0.21$ & $0.03$ & $4.0$ & $ { 3.7 }^{ +1.4 }_{ -0.7 }$ &  0.26 (0.6) \\
& [360, 400] & $0$ & $0.84 \pm 0.20$ & $0.02$ & $3.0$ & $ { 3.4 }^{ +1.5 }_{ -0.4 }$ &  0.50 (0.0) \\
& [400, 440] & $1$ & $0.64 \pm 0.18$ & $0.03$ & $4.2$ & $ { 3.2 }^{ +1.1 }_{ -0.1 }$ &  0.17 (1.0) \\
& [440, 580] & $2$ & $2.0 \pm 0.4$ & $0.03$ & $4.7$ & $ { 4.6 }^{ +1.5 }_{ -1.1 }$ &  0.50 (0.0) \\
& >580 & $1$ & $2.3 \pm 0.4$ & $0.03$ & $4.0$ & $ { 4.7 }^{ +1.7 }_{ -0.8 }$ &  0.50 (0.0) \\
 
\noalign{\smallskip}\hline\noalign{\smallskip}
\multirow{16}{*}{\rotatebox[origin=r]{90}{\SRThree}}  & ~~[90, 110] & $0$ & $1.08 \pm 0.18$ & $0.03$ & $4.8$ & $ { 3.8 }^{ +1.3 }_{ -0.5 }$ &  0.50 (0.0) \\
& [110, 130] & $5$ & $2.8 \pm 0.4$ & $0.06$ & $7.9$ & $ { 5.6 }^{ +1.7 }_{ -0.9 }$ &  0.08 (1.4) \\
& [130, 150] & $5$ & $4.1 \pm 0.6$ & $0.05$ & $6.7$ & $ { 5.8 }^{ +2.0 }_{ -1.2 }$ &  0.26 (0.6) \\
& [150, 170] & $2$ & $4.0 \pm 0.6$ & $0.03$ & $4.1$ & $ { 5.3 }^{ +2.3 }_{ -1.2 }$ &  0.50 (0.0) \\
& [170, 190] & $3$ & $3.9 \pm 0.4$ & $0.03$ & $4.9$ & $ { 5.2 }^{ +2.4 }_{ -1.1 }$ &  0.50 (0.0) \\
& [190, 210] & $7$ & $3.7 \pm 0.6$ & $0.07$ & $9.1$ & $ { 6.1 }^{ +2.2 }_{ -1.9 }$ &  0.10 (1.3) \\
& [210, 230] & $6$ & $3.5 \pm 0.7$ & $0.06$ & $8.8$ & $ { 6.2 }^{ +1.8 }_{ -1.7 }$ &  0.08 (1.4) \\
& [230, 250] & $4$ & $3.3 \pm 0.6$ & $0.04$ & $6.1$ & $ { 5.3 }^{ +1.7 }_{ -1.2 }$ &  0.28 (0.6) \\
& [250, 270] & $3$ & $2.5 \pm 0.4$ & $0.04$ & $5.1$ & $ { 4.5 }^{ +1.8 }_{ -1.3 }$ &  0.36 (0.4) \\
& [270, 300] & $3$ & $3.7 \pm 0.4$ & $0.03$ & $4.8$ & $ { 5.4 }^{ +1.8 }_{ -1.6 }$ &  0.50 (0.0) \\
& [300, 330] & $3$ & $3.0 \pm 0.4$ & $0.04$ & $5.1$ & $ { 4.9 }^{ +1.8 }_{ -1.1 }$ &  0.50 (0.0) \\
& [330, 360] & $2$ & $2.06 \pm 0.35$ & $0.03$ & $4.8$ & $ { 4.5 }^{ +1.5 }_{ -1.4 }$ &  0.50 (0.0) \\
& [360, 400] & $3$ & $3.2 \pm 0.7$ & $0.04$ & $5.1$ & $ { 5.3 }^{ +2.0 }_{ -1.5 }$ &  0.50 (0.0) \\
& [400, 440] & $0$ & $1.70 \pm 0.27$ & $0.02$ & $3.0$ & $ { 3.7 }^{ +2.0 }_{ -0.6 }$ &  0.50 (0.0) \\
& [440, 580] & $7$ & $4.3 \pm 0.5$ & $0.06$ & $8.7$ & $ { 6.2 }^{ +1.9 }_{ -1.3 }$ &  0.09 (1.3) \\
& >580 & $8$ & $4.6 \pm 0.6$ & $0.07$ & $10.0$~~ & $ { 6.5 }^{ +2.3 }_{ -1.4 }$ &  0.08 (1.4) \\
 
\noalign{\smallskip}\hline\noalign{\smallskip}
\end{tabular*}
}
\end{table}

The model-independent limits are summarized in Table~\ref{tab:discovery}, which includes for each signal region:
\begin{itemize}
\item the number of observed events \nObs,
\item the expected number of SM events \nExp and the associated uncertainty from a fit to the CRs only,
\item the observed limit on the visible cross section \visXS of the potential BSM process,
\item the corresponding observed upper limit on the number of BSM events \SObs,
\item the expected upper limit on the number of BSM events \SExp and the associated uncertainty,
\item and the $p$-value (and associated significance $Z$) for the SM background alone to fluctuate to at least the number of observed events.
\end{itemize}
The observed limit \visXS is defined as the ratio of \SObs to the integrated luminosity, and it incorporates the cross section, acceptance, and selection efficiency of the generic BSM signal.
No \mZl bin shows a significant excess in all three SRs, in contrast to what
would be expected in the presence of a resonance that contributes to all SRs.The largest excess of data over the expected background is seen in \SRTL for the \mZl region between 150 and 170~\GeV, with an associated significance of $2.1\,\sigma$.
This is consistent with the expectation from statistical fluctuations of the SM background when considering 48 independent signal regions.
 
\FloatBarrier
 
\subsection{Mass limits on \BL RPV production}
\label{sec:search:exclusion}
 
Hypothesis tests for the \BL signal models are performed using the same \CLs prescription~\cite{Read:2002hq},
with exclusion lower limits set on the \chono masses for various scenarios of the \chono branching fractions using asymptotic formulas~\cite{Cowan:2010js}.
A profile likelihood fit is performed simultaneously to the CRs and all \mZl bins of the three SRs,
benefiting from the contribution of a signal model to a small number of \mZl bins coherently across \SRTL, \SRFour, and \SRThree.
The signal strength is represented by a single parameter of interest and coherently scales the signal yield across all regions.
 
The sensitivity to the signal models is dependent on the \chono branching fractions to each lepton and boson type,
and a scan is performed over various combinations.
The contributions from the \CCsignal and \CNsignal processes are treated together, and the \chono branching fractions are treated as fully correlated.
Four scenarios are considered for the \chono branching fractions to leptons: the scenario with equal branching fractions to $e$, $\mu$, and $\tau$-leptons
and the three scenarios with 100\% branching fractions to a single lepton type.
 
For each leptonic scenario, the \chono branching fractions to $W$, $Z$, and Higgs bosons are scanned at 10\% intervals.
A 0\% branching fraction to $Z$ bosons is not explored and is replaced by a 1\% branching fraction in the scans.
No significant difference in sensitivity is seen for the relative \chono branching fractions to $W$ or Higgs bosons,
with the sensitivity dominated by the branching fraction to $Z$ bosons, which produces the target trilepton resonances.
The three SRs contribute roughly equally to the overall sensitivity of the search, with a minor increase in sensitivity to Higgs boson decays from \SRTL
offset by a similar increase in sensitivity to $W$ boson decays from \SRFour.
 
The expected and observed mass-exclusion contours as a function of the \chono branching fraction to $Z$ bosons
are shown in Figure~\ref{fig:limits} for each of the four lepton-flavor scenarios.
The \chono branching fractions to $W$ and Higgs bosons are set to be equal here.
Limits are set for signal masses above 100~\GeV, and agreement within the uncertainties is seen between the observed and expected limits.
The observed limit is slightly weaker than the expected limit due to the minor excesses seen at low \mZl in \SRFour and in some high \mZl bins in \SRTL and \SRThree.
 
The observed mass exclusions are strongest when the \chono branching fraction to $Z$ bosons is largest,
reaching 1100~\GeV and 1050~\GeV for the $e$ and $\mu$ channels, respectively.
The limit is slightly reduced to 975~\GeV when no assumption is made about the flavor of the directly produced lepton,
and is weakest at 625~\GeV when only \chono decays into $\tau$-leptons are allowed.
The observed mass limit becomes significantly reduced when the \chono branching fraction to $Z$ bosons falls below 20\%,
reaching 375~\GeV in the $\mu$ channel and 350~\GeV in the $e$ channel when the branching fraction reaches 1\%.
No limits are set when requiring decays into $\tau$-leptons for branching fractions to $Z$ bosons below 11\%.

\begin{figure}[h]
\centering
\includegraphics[width=0.46\textwidth]{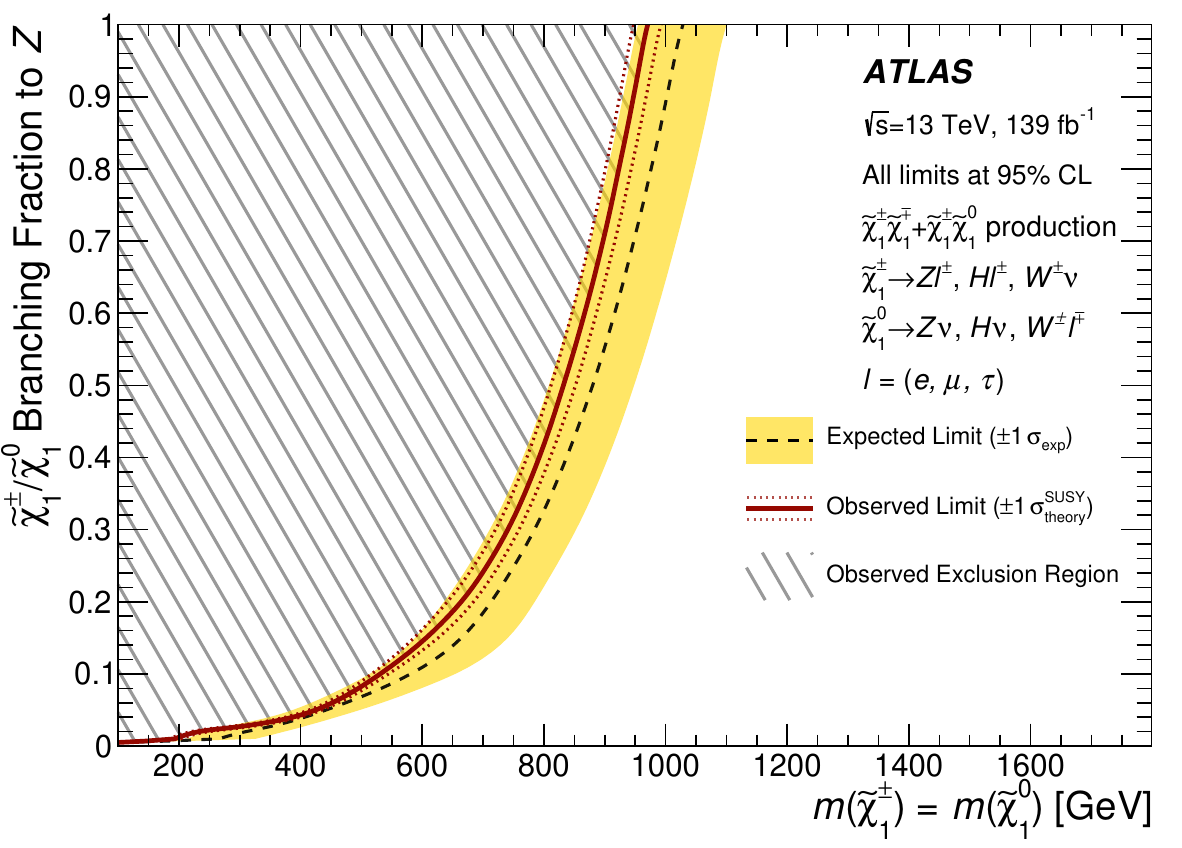}
\includegraphics[width=0.46\textwidth]{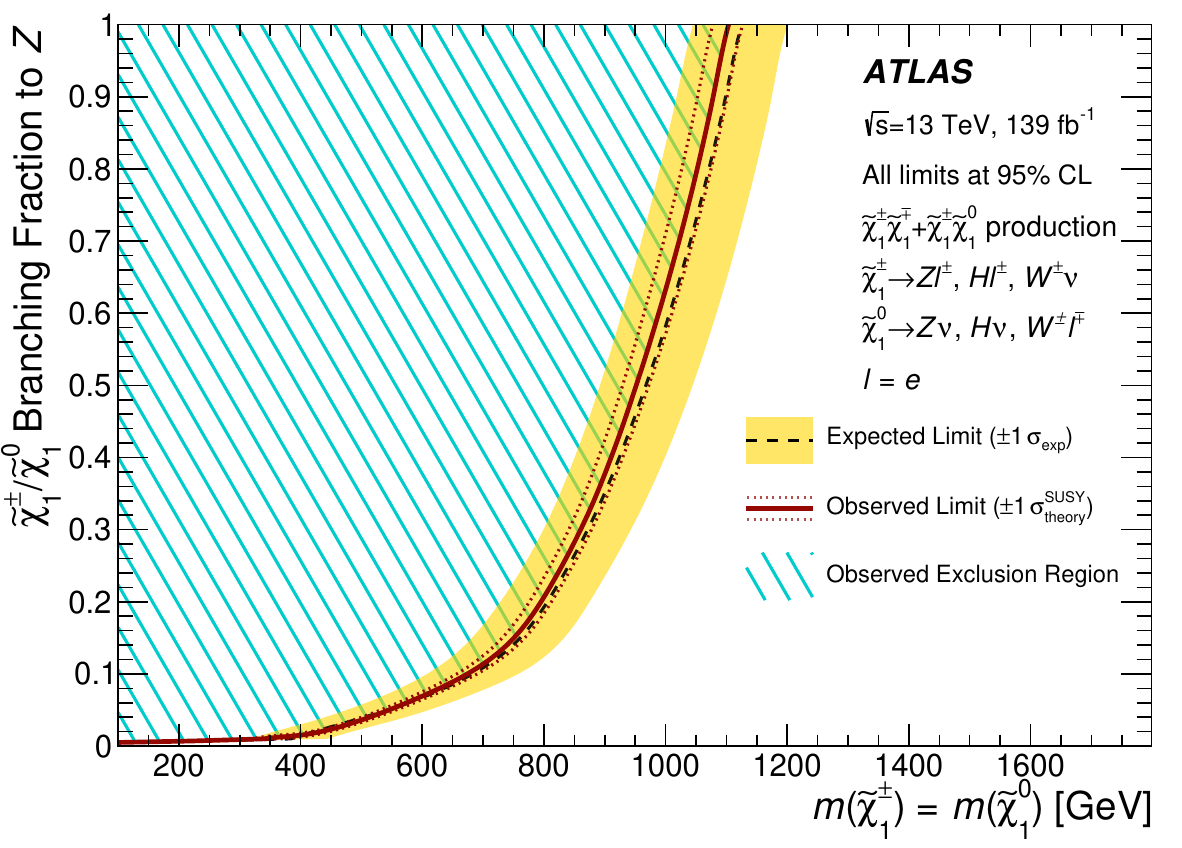}
\includegraphics[width=0.46\textwidth]{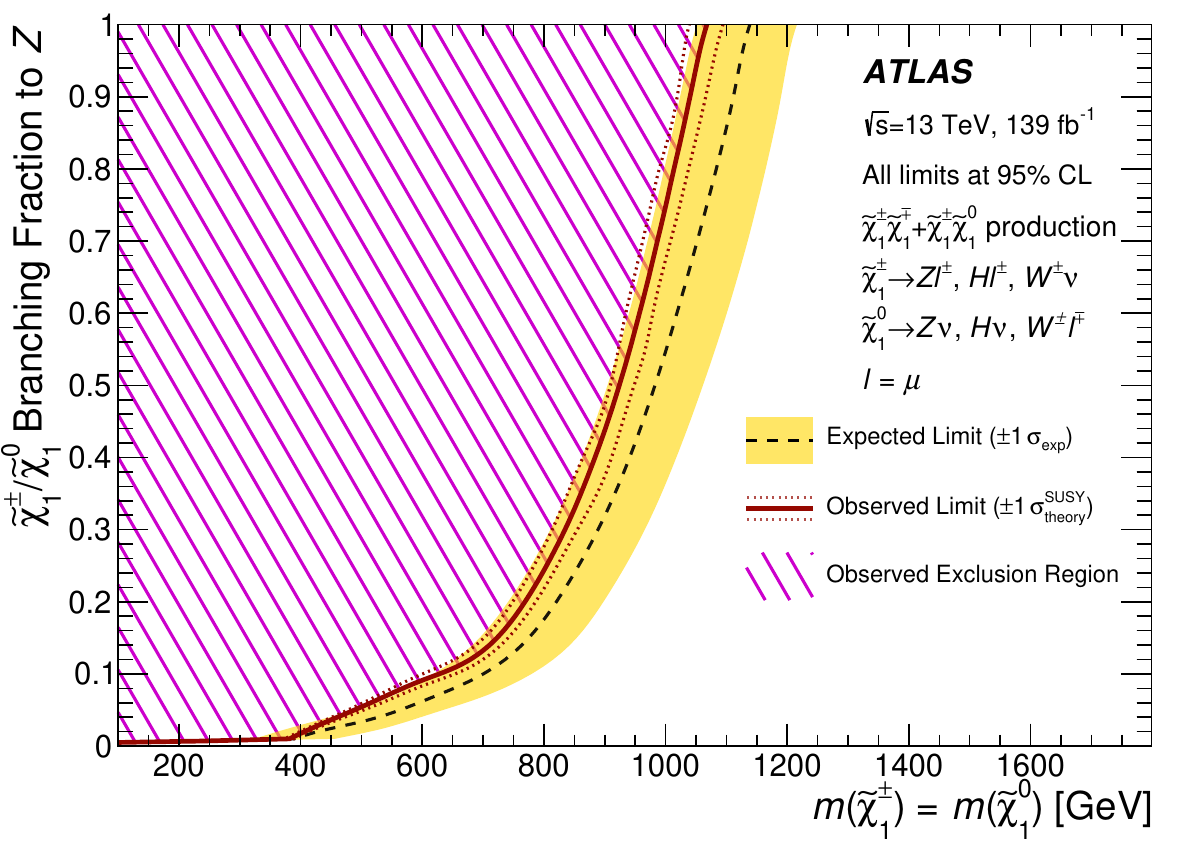}
\includegraphics[width=0.46\textwidth]{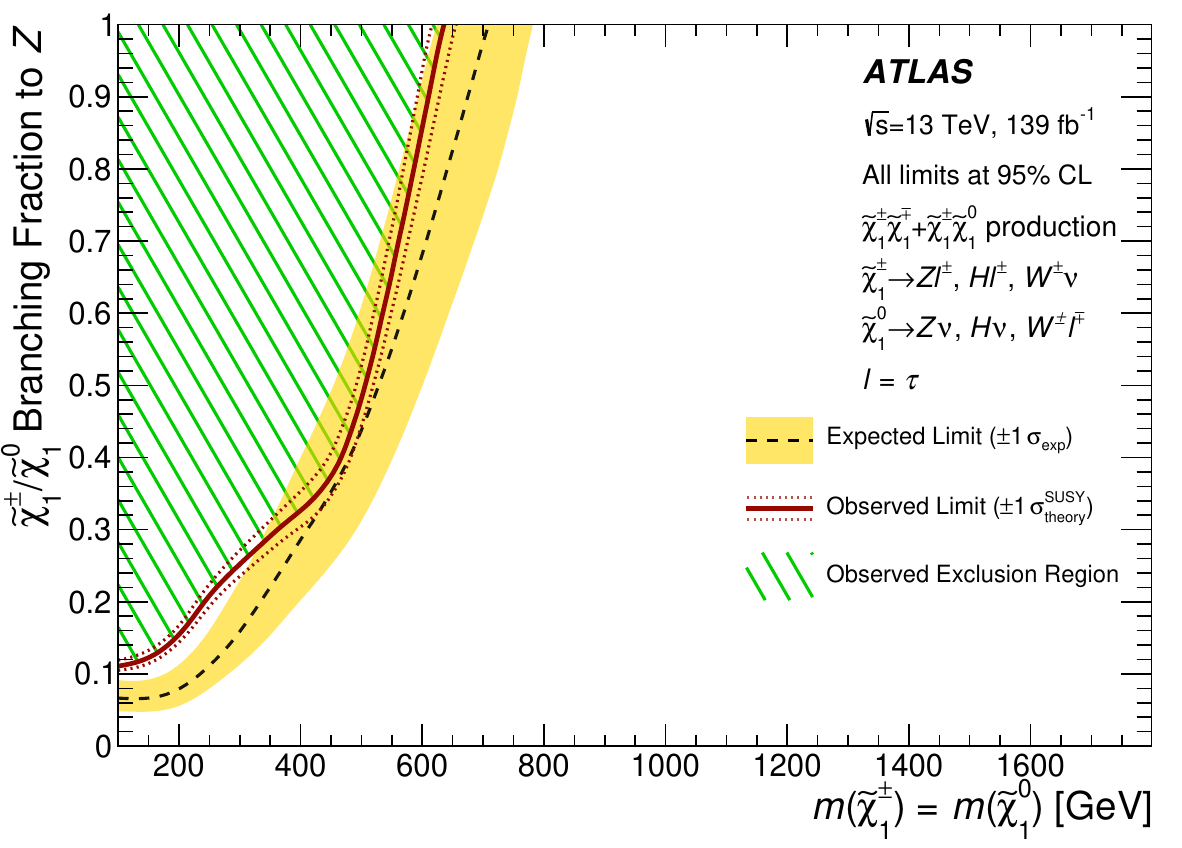}
 
\caption{ Exclusion curves for the simplified model of \CCsignal + \CNsignal production as a function of \chono mass and branching fraction to $Z$ bosons.
Curves are derived separately when requiring that the charged-lepton decays of \chono are into (top left) any leptons with equal probability,
(top right) electrons only, (bottom left) muons only, or (bottom right) $\tau$-leptons only.
\explimit
\obslimit
\exludedcolors
The sum of the \chono branching fractions to $W$, $Z$, and Higgs bosons is unity for each point, and the branching fractions to $W$ and Higgs bosons are chosen so as to be equal everywhere.
}
\label{fig:limits}
\end{figure}


\FloatBarrier

\section{Conclusions}
\label{sec:conclusion}
This paper presents a search for wino-type \CCsignal and \CNsignal production where
each \chono decays via an RPV coupling into a $W$, $Z$, or Higgs boson and a lepton.
The dataset corresponds to an integrated luminosity of $139~\ifb$ of proton--proton collision data produced at a
center-of-mass energy of $\sqs=13~\TeV$ and collected by the ATLAS experiment at the LHC between 2015 and 2018.
This search primarily targets the three-lepton decay of a \chone and is the first ATLAS analysis
using $\sqs=13~\TeV$ data to search for a resonance in the $m_{Z\ell}$ spectrum.
Three signal regions are defined that target events with three or more leptons and missing transverse momentum
or with two fully reconstructed \chono decays.
The observed event yields are found to be in agreement with Standard Model expectations, with no significant
excess seen in the \mZl distributions of the signal regions.
 
Model-independent limits are set at a 95\% confidence level for each \mZl bin in each signal region.
The largest excess of data over the expectation in the 48 model-independent regions is found to be $2.1\,\sigma$.
No trend is seen in the distribution of data excesses in \mZl bins across the three signal regions.
Model-specific lower limits are also set on the \chono masses for various decay branching fractions
into a lepton (electron, muon, or $\tau$-lepton) plus a boson ($W$, $Z$, or Higgs),
reflecting sensitivity to the neutrino mass hierarchy and the MSSM parameters of the \BL RPV theory.
For scenarios with large \chono branching fractions to $Z$ bosons, lower limits on the \chono masses are
set at 625~\GeV, 1050~\GeV, and 1100~\GeV for 100\% branching fractions to a $Z$ boson plus a
$\tau$-lepton, muon, or electron, respectively.
 

\section*{Acknowledgments}

We thank CERN for the very successful operation of the LHC, as well as the
support staff from our institutions without whom ATLAS could not be
operated efficiently.
 
We acknowledge the support of ANPCyT, Argentina; YerPhI, Armenia; ARC, Australia; BMWFW and FWF, Austria; ANAS, Azerbaijan; SSTC, Belarus; CNPq and FAPESP, Brazil; NSERC, NRC and CFI, Canada; CERN; ANID, Chile; CAS, MOST and NSFC, China; COLCIENCIAS, Colombia; MSMT CR, MPO CR and VSC CR, Czech Republic; DNRF and DNSRC, Denmark; IN2P3-CNRS and CEA-DRF/IRFU, France; SRNSFG, Georgia; BMBF, HGF and MPG, Germany; GSRT, Greece; RGC and Hong Kong SAR, China; ISF and Benoziyo Center, Israel; INFN, Italy; MEXT and JSPS, Japan; CNRST, Morocco; NWO, Netherlands; RCN, Norway; MNiSW and NCN, Poland; FCT, Portugal; MNE/IFA, Romania; JINR; MES of Russia and NRC KI, Russian Federation; MESTD, Serbia; MSSR, Slovakia; ARRS and MIZ\v{S}, Slovenia; DST/NRF, South Africa; MICINN, Spain; SRC and Wallenberg Foundation, Sweden; SERI, SNSF and Cantons of Bern and Geneva, Switzerland; MOST, Taiwan; TAEK, Turkey; STFC, United Kingdom; DOE and NSF, United States of America. In addition, individual groups and members have received support from BCKDF, CANARIE, Compute Canada, CRC and IVADO, Canada; Beijing Municipal Science \& Technology Commission, China; COST, ERC, ERDF, Horizon 2020 and Marie Sk{\l}odowska-Curie Actions, European Union; Investissements d'Avenir Labex, Investissements d'Avenir Idex and ANR, France; DFG and AvH Foundation, Germany; Herakleitos, Thales and Aristeia programmes co-financed by EU-ESF and the Greek NSRF, Greece; BSF-NSF and GIF, Israel; La Caixa Banking Foundation, CERCA Programme Generalitat de Catalunya and PROMETEO and GenT Programmes Generalitat Valenciana, Spain; G\"{o}ran Gustafssons Stiftelse, Sweden; The Royal Society and Leverhulme Trust, United Kingdom.
 
The crucial computing support from all WLCG partners is acknowledged gratefully, in particular from CERN, the ATLAS Tier-1 facilities at TRIUMF (Canada), NDGF (Denmark, Norway, Sweden), CC-IN2P3 (France), KIT/GridKA (Germany), INFN-CNAF (Italy), NL-T1 (Netherlands), PIC (Spain), ASGC (Taiwan), RAL (UK) and BNL (USA), the Tier-2 facilities worldwide and large non-WLCG resource providers. Major contributors of computing resources are listed in Ref.~\cite{ATL-SOFT-PUB-2020-001}.
 

\clearpage
\printbibliography
 
\clearpage 
 
\begin{flushleft}
\hypersetup{urlcolor=black}
{\Large The ATLAS Collaboration}

\bigskip

\AtlasOrcid[0000-0002-6665-4934]{G.~Aad}$^\textrm{\scriptsize 102}$,    
\AtlasOrcid[0000-0002-5888-2734]{B.~Abbott}$^\textrm{\scriptsize 128}$,    
\AtlasOrcid[0000-0002-7248-3203]{D.C.~Abbott}$^\textrm{\scriptsize 103}$,    
\AtlasOrcid[0000-0002-2788-3822]{A.~Abed~Abud}$^\textrm{\scriptsize 36}$,    
\AtlasOrcid[0000-0002-1002-1652]{K.~Abeling}$^\textrm{\scriptsize 53}$,    
\AtlasOrcid[0000-0002-2987-4006]{D.K.~Abhayasinghe}$^\textrm{\scriptsize 94}$,    
\AtlasOrcid[0000-0002-8496-9294]{S.H.~Abidi}$^\textrm{\scriptsize 167}$,    
\AtlasOrcid[0000-0002-8279-9324]{O.S.~AbouZeid}$^\textrm{\scriptsize 40}$,    
\AtlasOrcid{N.L.~Abraham}$^\textrm{\scriptsize 156}$,    
\AtlasOrcid[0000-0001-5329-6640]{H.~Abramowicz}$^\textrm{\scriptsize 161}$
% ...
\bigskip
\\

% ...
$^{36}$CERN, Geneva; Switzerland.\\
% ...
$^{40}$Niels Bohr Institute, University of Copenhagen, Copenhagen; Denmark.\\
% ...
$^{53}$II. Physikalisches Institut, Georg-August-Universit\"{a}t G\"ottingen, G\"ottingen; Germany.\\
% ...
$^{94}$Department of Physics, Royal Holloway University of London, Egham; United Kingdom.\\
% ...
$^{102}$CPPM, Aix-Marseille Universit\'e, CNRS/IN2P3, Marseille; France.\\
$^{103}$Department of Physics, University of Massachusetts, Amherst MA; United States of America.\\
% ...
$^{128}$Homer L. Dodge Department of Physics and Astronomy, University of Oklahoma, Norman OK; United States of America.\\
% ...
$^{156}$Department of Physics and Astronomy, University of Sussex, Brighton; United Kingdom.\\
% ...
$^{161}$Raymond and Beverly Sackler School of Physics and Astronomy, Tel Aviv University, Tel Aviv; Israel.\\
% ...
$^{167}$Department of Physics, University of Toronto, Toronto ON; Canada.\\
% ...

\end{flushleft}


\end{document}